\documentclass[12pt]{article}
\usepackage{natbib,amssymb,amsmath}
\usepackage{amsfonts}
\usepackage{graphicx}
\usepackage{amsmath}
\usepackage{amssymb}
\usepackage{color}
\usepackage{lineno}
\usepackage{dirtytalk}

\usepackage{comment}

\newcommand{\be}{\begin{equation}}
\newcommand{\ee}{\end{equation}}
\newcommand{\bea}{\begin{eqnarray}}
\newcommand{\eea}{\end{eqnarray}}
\newcommand{\ben}{\begin{equation*}}
\newcommand{\een}{\end{equation*}}

\newcommand{\eps}{\epsilon}
\newcommand{\hz}{\hat{\zeta}}
\newcommand{\hf}{\hat{f}}
\newcommand{\hg}{\hat{g}}
\newcommand{\ma}{\mathcal}

\begin{document}

\title {Evolution of wind-generated shallow water waves in a Benney-Luke equation}

\author {Montri Maleewong$^1$ and Roger Grimshaw$^2$}
\date{}
\maketitle
\vspace{-5ex}
\noindent
\begin{center}
$^1$ Department of Mathematics, Faculty of Science, Kasetsart University, \\ Bangkok, 10900, Thailand \\
$^2$ Department of Mathematics, University College London, London, WC1E 6BT, UK. 	
\end{center}

\begin{abstract}
 In  recent papers, denoted by  MG24, MG25 in this text, we used the Korteweg-de Vries (KdV) equation 
 and its two-dimensional extension, the  Kadomtsev-Petviashvili (KP) equation to describe the evolution 
 of  wind-driven water wave packets in shallow water.  Both equations were modified to include the effect of wind forcing, modelled using the Miles critical level instability theory.  In this paper that is extended to a Benney-Luke (BL) equation, similarly modified for wind forcing.  The motivation is that the BL equation is isotropic in the horizontal space variables, unlike the KP model, and noting that the KdV model is one-dimensional. 
The modified BL equation is studied using wave modulation theory as in MG24, MG25, and 
 with comprehensive numerical simulations. Despite the very different spatial structure the results 
 show that under the right initial conditions and parameter settings, solitary wave trains emerge 
 as in MG24, MG25.
 \end{abstract} 

\section{Introduction}

The  evolution of wind-generated water waves  is a highly studied topic, for the overlapping 
aims of scientific research and operational forecasting, see \citet{ja04, OS10, ww18} for instance.
For the most part  the emphasis has been on deep-water waves for the valid reason that in 
the  ocean and in many experimental wave flumes, wind waves do not feel the water depth.
Nevertheless it is somewhat surprising that studies of wind-driven waves in shallow water are 
relatively sparse until quite recently, see \citet{os14, zf21, ml23} and our recent work
\citet{mg24b}. Not surprisingly these were based on the Korteweg-de Vries (KdV) equation, 
which is well established as a model for shallow water waves, and widely known as a 
canonical integrable nonlinear wave equation. The inclusion of wind forcing terms, especially 
the crtical layer mechanism of \citet{mi57} leads to an (unstable) Burgers type of term
and we call that outcome a KdVB equation. Since the KdV equation is uni-directional, 
with only one horizontal space coordinate, $x$, in \citet{mg25} we examined a similar forcing 
extension  to the Kadomtsev-Petviashvili (KP) equation which, while still uni-directional,  
has a weak dependence on the spatial variable $y$, transverse to the wind direction ($x$); 
that outcome we call the KPB equation. Both the KdV and KPB models exhibit the
formation  of many solitary waves, a soliton gas, see the review by \citet{el21}. 

The KP equation is not isotropic in the spatial variables $x, y$ as the equation contains a 
scaling $y \sim x^2 $, unlike the full Euler equations which are isotropic in $x, y $.
This feature was remedied by \citet{bl64} who derived the Benney-Luke (BL)  equation 
for shallow water waves, which unlike the KP equation is isotropic in $x, y$. 
Here we add the same wind forcing term to the BL equation, which is again a Burgers
 type of term.  As shown by \citet{bl64}, see also  \citet{ac11}, the BL equation has the 
 same type of travelling waves, cnoidal and solitary, as the KdV and KP equations. 
 
 The BL equation is an asymptotic reduction of the full Euler equations based on a balance two 
 small parameters, $\eps $ measuring wave amplitude, and  $\beta $ measuring long wave dispersion.
 At leading order it is the two-dimensional linear wave equation, and importantly has two small 
 perturbative terms, weak nonlinearity and slow long-wave linear  dispersion, made equal with 
 $\beta = \eps $.  Consequently, initially the solution is close to that of the linear long wave  equation, 
 and the appearance of nonlinear and dispersive effects, such as the formation of solitary waves, 
 takes  a very long time, and occurs in a reference frame moving with the linear long wave speed,
 requiring very long-time numerical simulations. Importantly here the dispersion term is regularised
 to remove spurious high wavenumber instability generated by the asymptotic reduction, see
 \citet{whitham74, ehs16} for instance.
 
 In the next section \ref{bleq}, we present the asymptotic derivation of the forced regularised 
 BL equation. Then in section \ref{tw} we describe the cnoidal and solitary wave solutions, \ref{cn},
 followed by  their modulation theory, \ref{mod}, and in particular the modulation theory for a solitary 
 wave train, \ref{swt}. Our numerical simulations of the BL equation is described in section \ref{num}. 
 We conclude with a summary and discussion in section \ref{summary}. 
 
\section{Benney-Luke equation}
\label{bleq}

\citet{bl64} derived the BL equation as a  shallow-water, long-wave asymptotic reduction 
of the full Euler equations.  Here we outline that asymptotic theory, following the original derivation 
and adding to it wind forcing due to the critical level instability theory mechanism of \citet{mi57},
as introduced by \citet{mg24a, mg24b} into the full Euler equations. The water is an 
 inviscid, incompressible fluid of constant density in the  domain $-h < z< \zeta $ 
  where $z = -h$ is a rigid fixed bottom and $z =\zeta (x, y, t)$ is the free surface.  
For irrotational flow the governing equations are expressed in terms of a 
 velocity potential $\phi (x, y, z, t)$ so that the fluid velocity ${\bf u} = \nabla \phi$. 
\be\label{laplace}
\nabla^{2} \phi +\phi_{zz}  = 0\,, \quad \nabla^2  \phi = \phi_{xx}  + \phi_{yy}\,, \quad -h< z < \zeta  \,. 
\ee
The bottom boundary condition is 
\be\label{bottom}
\phi_z \, = \, 0  \,, \quad  \hbox{on} \quad  z = -h  \,. 
\ee
  At the free surface the kinematic and dynamic boundary conditions are
\be\label{kin}
\zeta_t  + \phi_x \zeta_x  +\phi_y\zeta_y - \phi_z = 0\,, \quad z = \zeta \,, 
\ee 
\be\label{dyn}
\phi_t + \frac {1}{2} (\phi_x^2 + \phi_y^2 + \phi_z^2) + g\zeta   =  -\frac{P_a}{\rho_w} 
\,,\quad z = \zeta \,.
\ee 
Here $\rho_w $ is the constant water density, and $P_a (x, y, t)$ is the surface air 
pressure containing the wind forcing term to be specified later. 
Next, we introduce scaled variables, denoted by   $[\bar{\cdot}]$,  and two 
dimensionless small parameters $\eps, \beta $,
\be\label{scale}
\begin{split}
&\phi = CL \beta \bar{\phi}\,,\,\, \zeta =A_0 \bar{\zeta}\,, \,\, (x, y) = L(\bar{x}, \bar{y}), \,\, z = h\bar{z}, 
\,\, t = \frac{L}{C}\bar{t}\,, \,\, C =(gh)^{1/2} \,,\\
& \qquad  \eps = \frac{h^2}{L^2} \,, \quad \beta = \frac{A_0}{h} \,, \quad 
\frac{P_a }{\rho_w} = \beta C^2 \bar{P_a} \,.
\end{split}
\ee
where $A_0, L$ are measures of wave amplitude and horizontal wave length scale respectively.

The equations (\ref{laplace} - \ref{dyn})  become, omitting the $[\bar{\cdot}]$
and putting $\beta = \eps $ as in \citet{bl64},
\bea
&\eps \nabla^2  \phi +\phi_{zz} =0\,, \quad -1< z < \eps \zeta  \,, \label{laplace1} \\
& \phi_z =0, \quad z=-1\,, \label{bottom1} \\
  & \eps \zeta_t  +  \eps^2 (\phi_x \zeta_x  +\phi_y\zeta_y )- \phi_z = 0\,, \quad 
  z = \eps \zeta \,, \label{kin1}\\
  & \phi_t + \frac {1}{2}(\eps (\phi_x^2 + \phi_y^2 )+ \phi_z^2) + \zeta  = 
  - P_a \,, \quad  z = \eps\zeta \,. \label{dyn1}
\eea

An asymptotic solution of (\ref{laplace1}, \ref{bottom1})  for $\eps \ll 1$ is
 \be\label{feq}
 \phi = f(x, y, t)  - \eps \nabla^2 f \frac{(z+1)^2}{2} + \eps^2 \nabla^4 f \frac{(z+1)^4}{24} +O(\eps^3)\,.
 \ee
 Then (\ref{dyn1}) yields
 \be\label{zf}
 \zeta = -f_t -\frac {\eps}{2}(f_x^2 + f_y^2  - \nabla^2 f_t)  - P_a .
 \ee
 The next step is to use equations (\ref{feq}, \ref{zf})  in ({\ref{kin1}) to get the
 regularised Benney-Luke equation,
 \be\label{bl}
f_{tt} - \nabla^2 f +\eps [-\frac{1}{3}  \nabla^2 f_{tt} +f_t \nabla^2 f  + 2f_x f_{xt} +  2f_y f_{yt}] + P_{at} =  0  \,. \\ 
 \ee
 This agrees with \citet{bl64} after using the leading order term to regularise and omitting $P_a $.
 
 We assume that the wind forcing acts in the $x$-direction, so that, following \citet{mi57}, 
 see also \citet{mg24b, mg25} and the references therein, the air pressure  is linked to the 
 surface slope; in non-dimensional variables,
 \be\label{pressair}
 P_{a} = \alpha \zeta_x, \quad \alpha = \eps^{1/2} \beta^{\prime}, 
 \quad \beta^{\prime }  = \beta_M \frac{\rho_a}{\rho_w} \frac{W_r^2 }{C^2 } \,.
 \ee
 $\rho_{a, w} $ are the air, water densities, $\beta_M$ is a dimensionless parameter 
 introduced by \citet{mi57}, and $W_r $ is a reference velocity.   As anticipated $P_a$ 
 is a small perturbation because $\rho_a \ll \rho_w $ and $W_r \ll C $. It can now be checked from  (\ref{bl})
 that $\beta_M \gtrless 0$ leads to instability or decay  for a periodic wave.\\
 
  Equation (\ref{bl}) contains a small parameter $\eps $ and initially to the lowest order in $\eps $, 
  the solutions are those of the linear long wave equation. Nevertheless, since it is the role of the 
  $O(\eps )$ terms which is the main interest, long-time simulations are needed. 
  A suitable value of $\eps $ is in the range $0.1 - 0.5 $, large enough to explore nonlinear and 
  dispersive effects, but well below large amplitude dynamical wave instability.  
  The BL equation (\ref{bl}) has an asymptotic reduction for $\eps \to 0$ to the KP equation 
  expressed in the variables  $X = x-t, Y = \eps^{1/2}y, T = \eps t $
  see \citet{ac11, ac13} and below.
    
  For the numerical simulations, the BL equation (\ref{bl}) is written as  a coupled system of two 
  first order in time equations for $f, g = -f_t $,  
   \be\label{bls}
   f_t = - g, \quad g_t = -\nabla^2 f  + \eps [\frac{1}{3}\nabla^2 g_{t}  - g\nabla^2 f  - 2f_x g_{x} -  2f_y g_{y}]
+  \alpha g_{xt} \,. \\ 
 \ee
 The surface displacement (\ref{zf}) is then given by
  \be\label{zfg}
 \zeta = g -\frac {\eps}{2}(f_x^2 + f_y^2  +\nabla^2 g)  -\alpha g_{x} \,.
 \ee
  The BL equation can also be converted to a conventional Boussinesq-type system for 
  $\zeta, U = f_x, V = f_y $, where $U, V$ are the leading order horizontal velocity field, 
  but we will not give details here.

The BL equation (\ref{bl}) is isotropic in  $(x,y)$  unlike the KPB equation of \citet{mg25}  in which
  $y \sim x^2 $ in scaling. As shown by \citet{ac11, ac13} and mentioned above, the BL equation 
  has an asymptotic reduction to the KPB equation as $\eps \to 0$, expressed to the leading order 
  in $\eps $ for a wave moving in the positive $x$ direction with a phase speed approximately unity, 
 \be\label{kpb}
(2\zeta_T +\frac{1}{3} \zeta_{XXX} + 3\,\zeta\,\zeta_X )_X + \zeta_{YY} + 
\frac{\alpha }{\eps} \zeta_{XXX} =0.
 \quad X = x-t, Y= \eps^{1/2}y, T =\eps t .
 \ee
  With some numerical  rescaling  this is the  KPB equation of \citet{mg25}.\\
  
   The initial conditions  are specification of $f= f_0(x,y), g = g_0(x,y)$. In general these can be 
   specified  independently, but here we will link them to assist in the generation of  waves 
   moving in the  positive $x$-direction. Two main cases A, B are considered,
   \be\label{ic}
   \begin{split}
  & A: \,\,  f_{0}(x,y) = \frac{A}{\omega_0 }\sin{( k_0 x+ l_0 y)}, \,\, g_{0}(x,y) = A\, \cos{( k_0 x+ l_0 y)},  \\
  & B: \,\, f_{0}(x,y) = \frac{A}{\omega_0 }\sin{( k_0 x)}\cos{( l_0 y)}, \,\,  g_{0}(x,y) = A\, \cos{( k_0 x)}\cos{( l_0 y)}, \\
  & \omega_0 =\frac{\kappa_0 }{(1 +\eps \kappa_0^2/3)^{1/2}}, \,\, \kappa_0  = (k_0^2 + l_0^2)^{1/2}. 
  \end{split}
\ee    
  $(f_{0}, g_{0})(x,y)$  are sinusoidal functions and $\omega_0 $ is the linear frequency for a wave 
  in the positive $k_0, l_0 $ direction.  $ A $ is the initial wave amplitude, and $k_0, l_0$ are the 
  initial wavenumbers, in case A generating a wave initially travelling in the $k_0, l_0$  direction.
   In case B, $g_{0}(x,y)$  is the initial elevation used  in the KPB simulation of \citet{mg25},
   but here in the BL equation (\ref{bls}) it will generate waves in both positive and negative $x$-directions. 
   We choose $f_0 (x, y)$ to correspond to linear waves moving in the positive $x$-direction 
   and use an upstream sponge layer to absorb waves moving in the negative $x$-direction.
   These initial conditions are designed  to produce wave packets, with many solitary waves, 
   resembling a soliton gas as found in the KdVB and KPB simulations of \citet{mg24b, mg25}.  
   As time increases  the nonlinear terms in (\ref{bls}) come into play, generating  solitary waves as 
   the dispersive and nonlinear effects come into balance.  The numerical simulations are on a 
   periodic domain $[-P_X <  x < P_X, \, -P_Y <  y < P_Y ]$.  The initial condition (\ref{ic}) is  
   combined with envelopes $ENV(x), ENV(y)$ with long length  scales $L_{X, Y}$ 
    such that $2\pi /k_0 \ll L_X \ll  P_X $, $2\pi /l_0 \ll  L_Y \ll P_Y $.  We usually ensure that
    $\omega_0 \sim 1 $ and vary  $k_0, l_0$ with typically $k_0 \sim 1$ and $ 0.1 \le l_0 \le 0.3  $.  
    The envelopes  $ENV(x), ENV(y) $ are chosen so that $f, g =0 $ at the periodic domain 
    boundaries,  see section \ref{num}. \\

   The mass is $\zeta $ and mass conservation is expressed in non-dimensional variables by (\ref{kin1})
   expressed here in the BL asymptotic form,
     \be\label{masseq}
    \zeta_t = -\nabla^2 f  - \eps [\frac{1}{6}\nabla^2 g_{t}  + g\nabla^2 f  + f_x g_{x}+ f_y g_{y}  ] \,.
     \ee
  In the full Euler system the energy equation is 
   \be\label{energydef}
   \begin{split}
   & \it{E}_t + \it{F}^{X}_x   + \it{F}^{Y}_y = - \alpha \zeta_x \zeta_t, \\
    & \it{E} =\frac{1}{2}\{\int_{-1}^{\eps \zeta} \,[\phi_x^2 +\phi_y^2  + \frac{\phi_z^2 }{\eps }]\,dz + \zeta^2 \}\,\\
    &  \it{F}^{(X, Y)} = -\int^{\eps \zeta}_{-1}\, \phi_t (\phi_x, \phi_y)\, dz. 
    \end{split}
   \ee
  Then using the asymptotic expansion (\ref{feq}) this becomes,
   \be\label{energyeq}
   \begin{split}
  & \it{E} =\frac{1}{2}\{(f_x^2 + f_y^2 )(1 +\eps \zeta ) + \zeta^2 \} + 
   \eps \{\frac{(\nabla f)^2  }{2} + \frac{(\nabla f)(f_{x}^2 + f_{y}^2 )}{6} \}, \\
   &  \it{F}^{(X, Y)} = -(f_t f_{\chi })(1 + \eps \zeta) + \eps \frac{1}{6}(f_{\chi} \nabla^2 f_t
   + f_{t}\,\nabla^2 f_{\chi}),  \quad \chi =(x, y).
  \end{split}
   \ee
   To leading order in $\eps $ the energy equation (\ref{energydef}) reduces to 
    \be\label{energyred}
    \begin{split}
    & \it{E}_t + (\zeta f_x )_x  +  (\zeta f_y )_y + \cdots =   -\alpha \zeta_t \zeta_{x},\\
    & \it{E} = \frac{1}{2}\{(f_x^2 + f_y^2 ) + \zeta^2 \} + \cdots, \quad \zeta = -f_t +\cdots.  
    \end{split}
    \ee
    This reduced form, which is obtained just from the linear long  wave equation, suffices for most purposes.
 
\section{Travelling waves}
\label{tw}

\subsection{Cnoidal waves}
\label{cn}

\citet{bl64} showed that the BL equation (\ref{bl}) has periodic cnoidal and solitary wave solutions,
 for which $f, g, \zeta $ depend on the phase variable $\vartheta = kx + ly -\omega t$. 
 The phase velocity normal to the wave front $\vartheta = \hbox{constant} $ is 
 $c = \omega/\kappa = 1 + \eps c_1 + O(\eps^2)$. Here, in order to consider modulated waves, 
 we extend that, and seek solutions of the form $f, g, \zeta $ which depend  on  $\theta( x, y, t )$  
  with wavenumbers $k, l $ and frequency $\omega $ given by
 \be\label{phase}
\theta_x = k, \,\,  \theta_y = l, \,\, -\theta_t  = \omega = \kappa c, \,\, 
\kappa = (k^2 + l^2)^{1/2}, \quad c = 1 +\eps  c_1 + O(\eps^2).
\ee
$k, l, c $ are slowly varying relative to the phase, but the scale for the slow variation is not connected to $\eps $.
In the sequel  omitted terms $[\cdots]$ include both higher order terms in $\eps $ and higher order 
slowly varying terms. The equations for conservation of waves come from eliminating 
$\theta $ from  (\ref{phase}), 
\be\label{conswave}
 k_t + \kappa_x + \eps (\kappa c_1 )_x  + \cdots=0,  \quad
  l_t + \kappa_y  + \eps (\kappa c_1 )_y + \cdots =0,  \quad k_y =l_x \,. 
\ee
To the leading order in $\eps $, (\ref{conswave}) are equations for $k, l $ alone;
the terms in $\eps c_1$ will contain a dependence on other wave parameters such 
as the wave amplitude. 

The modulated wave  $\zeta (\theta, x, y, t) $  is periodic in $\theta $
with period $ 2\pi$ and the slow dependence on $x, y, t $ will be  
expressed through the wave parameters. We define a wave average for a single wave
\be\label{waveav}
<[\cdots ]> =  \frac{1}{2\pi }\int_{-\pi}^{\pi}\, [\cdots]\, d\theta \,.
\ee
The wave is then decomposed into a periodic wave $\hz $ with  zero mean, and  a mean level $D$
  \be\label{modwave}
\zeta = \hz(\theta, x, y, t) + D (x, y, t)\,, \quad  <\hz>  =0, \quad D( x, y, t )= <\zeta > \,.
\ee
The explicit $x, y, t$ dependence in $\hz, D$ is slowly varying.  Wave energy is usefully defined as 
$ E = <\hz^2> $ obtained from  (\ref{energyred}).  Correspondingly, $f, g$ are also decomposed into a 
periodic wave $\hf, \hg $ with  zero means, and mean values $F, G$, 
\be\label{modfg}
 \hg  = \omega \hf_{\theta } -\hf_t , \quad G = -F_t \,.
\ee
The leading order horizontal velocity field $f_x, f_y $ is given by
\be\label{vel}
f_{x, y} = (k, l)\hf_{\theta } + \hf_{x,y} + F_{x,y} \,.
\ee  \\
Substitution of (\ref{modwave},  \ref{modfg})  into the BL equation (\ref{bls}) yields at the leading order
in both $\eps $ and with respect to slow variation in $x, y, t$,
\be\label{blwave}
-2c_1 \hz + \frac{\kappa^2}{3} \hz_{\theta \theta  } +  \frac{3}{2}\kappa \hz^2 =  \hbox{constant} \,.  
\ee
Equation (\ref{blwave}) has the same structure as that for travelling waves of the KdV equation, 
see \citet{mg24b} for instance,  and hence has a cnoidal wave solution,
\bea
& \qquad  \qquad \hz = A\{b + cn^2(\gamma \theta; m)\}, \quad b  = \frac{1-m}{m}-\frac{E(m)}{mK(m)}\,,\label{cn1} \\
& A = 3\, m \,\Gamma^2 \,, \,\, c_1 - D = \frac{2\Gamma^2 }{3}\left\{2-m -\frac{3E(m)}{K(m)}\right\}\,,  
\,\,  \Gamma = \gamma \kappa \,. \label{cn2}
\eea
 To leading order  the corresponding expressions for  $f, g$  are given by 
 $\omega  \hf_{\theta } =  \hg  =\hz$. 
 
 In (\ref{cn1}, \ref{cn2}) $cn(x; m)$ is the Jacobi elliptic function of modulus $m$ ($0<m<1$) 
and $K(m), E(m)$  are the elliptic integrals of the first and second kind respectively, all defined by
\be\label{cndef}
 cn(x;m) = \cos{(\phi)} \,, \quad x = \int_{0}^{\phi }\, \frac{d\phi^{\prime }}{(1-m\sin^2 {\phi^{\prime }})^{1/2}} \,, 
\quad 0 \le \phi \le \frac{\pi }{2} \,,
\ee
\be\label{K}
K(m) = \int_{0}^{\pi /2 }\, \frac{d\phi }{(1-m\sin^2 {\phi })^{1/2}} \,, \quad 
E(m) = \int_{0}^{\pi /2 }\, (1-m\sin^2 {\phi })^{1/2}\, d\phi  \,.
\ee
The cnoidal wave  (\ref{cn1}) is periodic in $\theta $ with a period $2\pi = 2K(m)/\gamma $, 
which defines $\gamma  = K(m)/\pi $. The corresponding spatial period  is  $2\pi/k  =2K(m)/\Gamma $.   
$A$ is the wave amplitude  and the maximum, minimum values of $\hz $  are $A_M = A(b + 1), A_m = Ab$. 

When the modulus $m \to 0$,  $cn(x; m) \to \cos{(x)}$, $\gamma \to 1/2 $, $ b \to -1/2 $
and then the cnoidal wave (\ref{cn1}) collapses to a linear sinusoidal wave,
\be\label{linwave} 
\hz = \frac{A}{2}\,\cos{( \theta)}\,, \quad  c_1 - D = -\frac{\kappa^2 }{6} \,.
\ee
In this limit $A \to 0$, but $A/m = 3\Gamma^2 $ stays finite.  This linear sinusoidal wave
has a linear dispersion relation $\omega^2 (1 + \eps \kappa^2 /3) = \kappa^2 $,
relative to the mean level $D$, which reduces to the above expression for $c_1 $
as $\eps \to 0$. {\it{Inter alia }} this demonstrates that the BL equation (\ref{bl}) is linearly stable.

When the modulus $m \to 1$, $cn(x; m) \to \hbox{sech}(x)$, $K(m) \to \infty$, $E(m) \to  1 $
 and $b(m) \to 0$.  The  cnoidal wave (\ref{cn1}) becomes a solitary wave train,  
 defined on a periodic lattice, and riding on a background level $D$. 
\be\label{soltrain}
\hz =   A\,\hbox{sech}^2 (\gamma \theta )\,, \quad  A = 3\Gamma^2  \,, \quad  
c_1 - D = \frac{2\Gamma^2 }{3}\,.
\ee
In this limit, $\kappa \to 0$, $\gamma \to \infty$ but $\Gamma = \gamma \kappa $ stays finite.
Note that $\gamma \theta =\Gamma \Theta $, $\Theta =\theta/\kappa $ and we can usefully write that 
$\Theta = kx/\kappa +ly/\kappa - ct$ since $\omega  =\kappa c$.

\subsection{Modulation theory}
\label{mod}

The cnoidal wave (\ref{cn1}) has seven parameters $[A, m, \Gamma, D, c, k, l]$. 
Three modulation equations are provided by the conservation of wave equations 
(\ref{phase}, \ref{conswave}); there are two relations in (\ref{cn2}); the remaining two 
modulation equations are provided by taking the wave average of  the conservation of 
mass equation (\ref{masseq}) and the conservation of energy equation 
(\ref{energydef}, \ref{energyeq}).  

The wave average   of the conservation of mass equation (\ref{masseq}) yields, to leading order, 
\be\label{massmod}
\begin{split}
& D  = G -\frac {\eps}{2}(F_x^2 + F_y^2)   - \eps E, \quad G = -F_t .\\
& D_t = -\nabla^2 F  - \eps ( G\nabla^2 F  + F_x G_{x}+ F_y G_{y})  - \eps E_t .
\end{split}
\ee
With only an order $\eps$ dependence on wave energy $E$ effectively this is an 
equation for $D$ alone, but we note that $D$  increases/decreases in the opposite sense to $E$.
Hence to within an order $\eps $ term, $D( x, y, t)$ can  be regarded as known.
 In the sequel, we shall usually put $D =0, G =0, F =0$.  
 
 Then, again to leading order in $\eps $,  the wave average of the conservation of energy
equation (\ref{energydef}, \ref{energyeq}), using the reduced form (\ref{energyred}), yields an equation  
for wave energy,
\bea
&E_t + (\frac{k}{\kappa} E)_x  +  (\frac{l}{\kappa} E)_y  + \cdots =  \alpha J  \,,\label{enmod} \\
& E = <\frac{1}{2}\{f_x^2 + f_y^2 + f_t^2 \} > = <\hz^2 >  =
A^2(C_4 - b^2)\,, \quad C_4 = <cn^4>  \,,  \label{Emod}\\
& J = -<\zeta_t \zeta_{x} > = \omega k <\hz_{\theta}^2 >  =  
 \frac{k}{\kappa} \frac{16A^2\Gamma^2}{3}\{ -(1-m)b  +(2m-1)C_4 -mC_6 \}\,.  \label{F}
\eea
Here $C_n = <cn^n >\,, n=2, 4, \ldots $ where $C_2 =-b$ and espressions for $C_4, \ldots $
can be found in \citet{gy16}. The wave energy propagates in the $x-y$ plane with a velocity 
$k/\kappa, l/\kappa $ which at  this leading order in $\eps $ is the phase velocity.
Since here $\omega =\kappa c \approx \kappa > 0$, $kF > 0$,  (\ref{enmod}) implies  wave growth as 
$\alpha > 0$ due a critical level instability in the wind shear profile. But we note that as shown by 
\citet{mi57, kkmt10, mg24a, mg24b}  sufficient boundary layer dissipation at the water surface
leads to a revised $\alpha < 0 $ and wave decay. 

In the linear sinusoidal limit $m \to 0$ (\ref{linwave}) the modulation equation (\ref{enmod})
 where (\ref{Emod}, \ref{F}) reduce to 
\be\label{linmod}
 E  = <\hz^2 > = \frac{A^2}{8}\,, \quad
F =  \omega k <\hz_{\theta}^2 >  =  \omega k =\frac{A^2}{8} \,. 
\ee

In the solitary wave limit $m \to 1$ (\ref{soltrain}), $\kappa \to 0$, $E \to 0$
but $\ma{E} = E/\kappa  $ stays finite.  Similarly $F \to 0$ but $\ma{F} = F/\kappa  $ 
stays finite and (\ref{enmod}) becomes 
\bea
& \ma{E}_t  + \frac{k}{\kappa}\ma{E}_x + \frac{l}{\kappa}\ma{E}_y  =\alpha \ma{F} \,, \label{solen} \\
 & \ma{E} =  \frac{2A^2}{3 \pi \Gamma } = \frac{6\Gamma^3 }{\pi }, \quad  \ma{F} =  
 \frac{k}{\kappa }\frac{32 A^2 \Gamma }{45 \pi }  = \frac{k}{\kappa } \frac{32 \Gamma^5  }{5 \pi } \,. \label{EF}
\eea

\subsection{Solitary wave train}
\label{swt}

The modulation equations for a solitary wave train are (\ref{solen}, \ref{EF}, \ref{conswave}) which can 
be expressed in terms of $\Gamma $, since $A = 3\Gamma^2 $ for a solitary wave (\ref{soltrain}),
\bea
& \Gamma_t  + \frac{k}{\kappa}\Gamma_x + \frac{l}{\kappa}\Gamma_y  =
\alpha \frac{k}{\kappa } \frac{16 \Gamma^3 }{45} \,, \label{Gmod} \\
& k_t + \kappa_x + \eps (2 \kappa \Gamma^2 )_x =0,  \quad
  l_t + \kappa_y  + \eps (2 \kappa \Gamma^2 )_y =0,  \quad k_y =l_x \,. \label{conssw}
\eea
The modulation variables have been reduced to  [$\Gamma, k, l $] where $k, l$ 
satisfy (\ref{conssw}), which to leading order in $\eps $ we remind are equations for $k, l $ alone.

When the modulation depends only on $t$, then $k, l $ are constants, and 
\be\label{Gt}
\Gamma^2 = \frac{\Gamma_0^2 }{1 - \alpha^{\prime} t} , \quad  
\alpha^{\prime } = \alpha \frac{k}{\kappa} \frac{32 \Gamma_0^2 }{45} \,.
\ee
where $\Gamma_0 = \Gamma (t=0)$ is the constant initial value. When $ \alpha > 0 $ 
there is wave growth, $\Gamma \to \infty $ in finite time, $t =1/\alpha^{\prime}$.
This apparent singularity is essentially the same as that in the KdVB equation 
as found by \citet{ml23, mg24b}, but it is a singularity in the modulation theory and not in 
the full equations. When $\alpha < 0$ the wave decays, $\Gamma \to 0$ as $t \to \infty $ 
on a time scale proportional to $|\alpha^{\prime} |^{-1}$.
 
 Next, we consider modulations only in $t, x$ where we recall that the wind forcing is in the $x$-direction. 
Then the $y$-wavenumber $l = l_0$ is constant, and the modulation equations 
(\ref{Gmod},  \ref{conssw}) reduce to 
\bea
& \Gamma_t  + S(k)\Gamma_x  = \alpha S(k) \frac{16 \Gamma^3 }{45 } \,, \label{Gmodx} \\
& k_t + S(k)k_x  + \eps (2 \kappa \Gamma^2 )_x =  0, \quad S(k) = \frac{k}{\kappa}, \quad 
\kappa  =(k^2 + l_0^2)^{1/2}  \,. \label{kx}
\eea
To leading order in $\eps $, equation (\ref{kx}) is a Hopf equation for $k$ alone, with implicit solution 
\be\label{ksol}
k = k_{(0)}(x - S(k)t),  \quad k(x, t=0) = k_{(0)}(x) \,.
\ee
This solution exists provided that $1 + k_{(0)x}S_k t > 0$, otherwise steepens and breaks; 
since $S_k > 0$, breaking will occur if $k_{(0)x} < 0$ somewhere.  With $k$ known, equation 
(\ref{Gmodx}) is a hyperbolic equation for $\Gamma $ alone. The solution propagates with speed 
$S(k)$ modulo growth or decay as $\alpha \gtrless 0$, similar to (\ref{Gt}),  
due to the term on the right hand side. 

When the $\eps $ term is retained in (\ref{kx}) equations (\ref{Gmodx}, \ref{kx})
 form two coupled nonlinear hyperbolic equations.  As a guide, we seek a 
 simple asymptotic solution with $k = k_0 + \eps k_1 +\cdots $ so that  (\ref{Gmodx}, \ref{kx}) become
 \bea
 & \Gamma_t  + S_0 \Gamma_x  = \alpha S_0 \frac{16 \Gamma^3 }{45 } \,, \quad S_0 = S(k_0) \,, \label{Gmodxe}  \\
 &  k_{1t} + S(k_0)k_{1x} + (2\kappa_0 \Gamma^2)_x  = 0 \,. \label{k1t}
 \eea
The general solution is readily found in  the variables  $X = x- S_0 t, t $. 
In the travelling frame  $\Gamma $ is given by (\ref{Gt}) with $\Gamma_0 =\Gamma_0(X)$. 
Then $k_{1t} = -4 \kappa_0  \Gamma \Gamma_X $;  since $\Gamma_x /\Gamma^3 
=  \Gamma_{0X} /\Gamma_{0}^3 $, $k_{1t}  = -4 \kappa_0 \Gamma^4 \Gamma_{0X}/\Gamma_{0}^3 $
and using (\ref{Gt}) the general solution  is readily found,
\be\label{k1}
k_1 =-4 \kappa_0 \Gamma_0 \Gamma_{0X} \frac{1}{\alpha^{\prime}(1 - \alpha^{\prime} t )} \,.
\ee
Hence  $k_1 $ decreases or increases according as  $\Gamma_{0X} \gtrless 0$.  
 When $\alpha > 0$, $\Gamma \to \infty $ in finite time and $k_1 \to -\infty $, 
 invalidating the approximation of $k$.  When $\alpha< 0$, $\Gamma \to 0$ and 
 $k_1  \to 0$ as $t \to \infty $

Analogous solutions can be found for modulations in $t, y$ only, essentially by interchanging 
$x, k $ with $y, l$.  Briefly, the $x$-wavenumber $k=k_0$ is a constant, 
and the modulation equations (\ref{Gmod},  \ref{conssw}) reduce to 
\bea
& \Gamma_t  + S^{*}(l)\Gamma_y  = \alpha \frac{k_0}{\kappa} \frac{16 \Gamma^3 }{45 } \,, \label{Gmody} \\
& l_t + S^{*}(l)l_y + \eps^2(2 \kappa  \Gamma^2 )_y  = ,\quad 
S^{*}(l) = \frac{l}{\kappa}, \quad \kappa  =(k_0^2 + l^2)^{1/2}  \,. \label{ly}
\eea
The slopes $ S(k), S^{*}(l) $ are perpendicular.  To leading order in $\eps $, 
equation (\ref{ly}) is a Hopf equation for $l$ alone, with implicit solution 
\be\label{lsol}
l= l_{[0]}(y - S^{*}(l)t),  \quad l(y, t=0) = l_{[0]} (y) \,.
\ee
Similarly to (\ref{ksol}) this solution exists provided that $1 + l_{[0]y}S^{*}_l t > 0$, otherwise steepens and breaks; 
since $S^{*}_l > 0$, breaking will occur  if $l_{[0]y} < 0$ somewhere.  With $l$ known, the solution equation 
(\ref{Gmody}) propagates with speed $S^{*}(l)$ modulo growth or decay as $\alpha \gtrless 0$, 
Retention of the term in $\eps  $ in (\ref{ly}) will lead to a similar  analysis  to that above 
leading to  (\ref{Gmodxe}, \ref{k1t}) for $x, t $ modulations, with similar findings. Briefly there is a	
simple asymptotic solution with $l = l_0 + \eps l_1 +\cdots $ so that  (\ref{Gmody}, \ref{ly}) become
\bea
 & \Gamma_t  + S^*_0 \Gamma_y  = \alpha S^*_0 \frac{16 \Gamma^3 }{45 } \,, 
 \quad S^*_0 = S^*(l_0) \,, \label{Gmodxe}  \\
 &  l_{1t} + S^*(l_0)l_{1y} + (2\kappa_0 \Gamma^2)_y  = 0 \,. \label{l1t}
 \eea
 As before the general solution is  found in  the variables  $Y= y- S^*_0 t, t $. 
$\Gamma $ is given by (\ref{Gt}) with $\Gamma_0 =\Gamma_0(y)$, and
then $l_{1t} = -4 \kappa_0  \Gamma \Gamma_Y $.  The general solution  of (\ref{l1t}) is 
\be\label{l1}
l_1 =-4 \kappa_0 \Gamma_0 \Gamma_{0Y} \frac{1}{\alpha^{\prime}(1 - \alpha^{\prime} t )} \,.
\ee
Hence  $l_1 $ decreases or increases according as  $\Gamma_{0Y} \gtrless 0$.  
 When $\alpha > 0$, $l_1 \to -\infty $, invalidating the approximation of $l$.  
 When $\alpha< 0$, $\Gamma \to 0$ and $l_1  \to 0$ as $t \to  \infty $.

\section{Numerical Results}
\label{num}

The BL equation written in the system form (\ref{bls}) is solved numerically
with a pseudo-spectral method in space and a fourth-order Runge-Kutta method in time.
A description of the code is in the Appendix.  Pseuodo-spectral methods are a popular 
choice for nonlinear wave equations since the pioneering work of \citet{fw78}, see
\citet{mt99, klein11, gkp18} for instance, and was used by us in a study of the 
 KdV-Burgers (KdVB) and KP-Burgers (KPB) equations, \citet{mg24b, mg25}. The code was tested 
using the known cnoidal and solitary wave solutions of the BL equation (\ref{bls}). 
Sponge layers analogous to those used in \citet{mg24b, mg25})  were sometimes used at the 
outer periodic boundaries to absorb outgoing waves. Details of the numerical scheme
 are described in the Appendix.

  In the numerical simulations, the initial conditions (\ref{ic}) are expressed on the periodic domain, 
  $[-P_X <  x < P_X, \, -P_Y <  y < P_Y ]$. 
  \be\label{icnum}
\begin{split}
& g_{0}(x,y) = ENV(x) ENV(y)  A\,COS(x, y), \,\, f_{0}(x,y)  = \frac{A}{\omega_0}\, ENV(x) ENV(y) SIN(x , y).\\
 & A: \quad  COS(x, y) =  \cos{( k_0 x +l_0 y)}, \quad SIN(x, y) =    \sin{( k_0 x+ l_0 y)}, \\
 &B:  \quad COS(x,y) =  \cos{( k_0 x)}\cos{( l_0 y)}, \quad SIN(x,y) = \sin{( k_0 x)}\cos{( l_0 y)}, \\
& ENV(x)  = \hbox{sech}(x/L_X) SMO(x)\,, \quad  ENV(y)  =   \hbox{sech}(y / L_Y) SMO(y), \\
& SMO(x) =  \exp{\{ - \frac{x^{2}}{(P_X^2 - x^2)}\}}\,, \quad 
SMO(y) =   \exp{\{-  \frac{y^{2}}{(P_Y^2 - y^2)}\}} \,.
\end{split}
\ee  
 $COS(x,y), SIN(x,y )$ are the sinusoidal functions in (\ref{ic}) of section \ref{bleq}, with wavenumbers 
 $k_0, l_0 $.  In the KPB simulations of \citet{mg25} we used an initial condition in the form of case B.
 This can decomposed as  $A( \cos{(k_0 x + l_0 y )} + \cos{(k_0 x  - l_0y )} ) /2 $,  each generating a 
 wave packet. In the KPB equation only the right-going wave is followed.  Hence in the present 
 simulations, in case A we choose an initial condition to generate a right-going wave, where 
 $f_0 = g_0 /\omega_0 $ based on the solution of the linearised BL equation. In case B we retain the
  initial condition $g_{0}(x,y)$  for the KPB simulations of \citet{mg25}, and adjust $f_0(x, y)$
  to favour a right-going wave as far as possible.  We usually set $\omega_0 \sim 1 $
  where also $\kappa_0 \sim 1 $, and then vary  $l_0$ over the range  $0 \le  l_0 \le 0.3  $, 
  so that $l_0 $ is of order $\eps^{1/2} $ for comparison with the  KPB equation (\ref{kpb}). 
 Any left-going wave is absorbed upstream  as if necessary we shift the initial wave location upstream, 
 so that the downstream wave can be followed for as long as possible.   As  anticipated  in section \ref{bleq}, 
 $ENV(x), ENV(y)$ are wave envelopes  with  length scales $\pi /k_0 \ll L_X \ll P_X$, $\pi /l_0 \ll  L_Y \ll P_Y $.
 When used, $SMO(x), SMO(y)$ are smoothing functions designed to ensure that $f_0, g_0 = 0$ 
 on and near the boundaries $x= \pm P_X, y = \pm P_Y $.  Otherwise, we put $SMO (x), SMO(y) =1$.
noting that in an infinite domain, $P_X, P_Y \to \infty $, $SMO(x), SMO(y) \equiv 1$.
  
The solutions of the BL system (\ref{bls}) depend on the parameter $\eps $, the forcing parameter $\alpha $, 
and the parameter set $[A, k_0 , l_0, L_{X}, L_{Y}]$ in the initial conditions (\ref{icnum}). 
The asymptotic parameter $\eps  $ is explicitly removed from (\ref{bls}) by the rescaling 
 \be\label{resc}
 (\hat{x}, \hat{y}, \hat{t}) =\eps^{-1/2} (x, y, t), \,\, \hat{f}= \eps^{1/2}\,f, \,\,\hat{g}= \eps g, \,\,
 \hat{\zeta} = \eps \zeta.
 \ee
  Then (\ref{bls}) becomes the  same equation but with $\eps =1$,  that is, after removing 
 the superscript $[\hat{\cdot}] $, 
   \be\label{bls1}
  \begin{split}
 &  f_t = - g, \quad g_t = -\nabla^2 f  +  [\frac{1}{3}\nabla^2 g_{t}  - g\nabla^2 f  - 2f_x g_{x} -  2f_y g_{y}]
+  \beta^{\prime}  g_{xt} \,,\\
 & \zeta = g -\frac {1}{2}(f_x^2 + f_y^2  +\nabla^2 g)  - \beta^{\prime} g_{x}, \quad
 \beta^{\prime} = \frac{\alpha }{\eps^{1/2}}\,. 
 \end{split}
 \ee
  The effect of the parameter $\eps $ can be recovered since if $f( x, y, t), g( x, y, t), \zeta (x, y, t)$ 
  is a solution of (\ref{bls1}) then $\eps^{-1/2}\,f(\eps^{-1/2} (x, y, t)) $, $\eps^{-1}\,g(\eps^{-1/2} (x, y, t))$, 
 $ \eps^{-1}\, \zeta (\eps^{-1/2} (x, y, t)) $ is a solution of (\ref{bls}).  However, the parameter $\eps $ 
 has an implicit effect through the initial conditions which we discuss below. 
 This rescaling is similar to an inversion of the original asymptotic reduction of \citet{bl64} 
 described in section \ref{bleq}, but it is not exactly equivalent to that;  it is used here 
 mainly as a numerical device, with the aim of creating a smaller time for nonlinear and 
 dispersive effects to appear, leading to solitary wave formation. Especially note that from (\ref{pressair}) 
  $\alpha /\eps^{1/2}  = \beta^{\prime} $ and does not depend on $\eps $. 
  Likewise, the KPB equation (\ref{kpb}) expressed in these rescaled variables  (\ref{resc}),
  is formally unchanged after  again omitting $[\hat{\cdot}]$,
\be\label{kpbn1}
(2\zeta_T +\frac{1}{3} \zeta_{XXX} + 3\,\zeta\,\zeta_X )_X + \zeta_{YY} + 
\beta^{\prime } \zeta_{XXX} =0; \,\, 
 X = x-t, Y= y, T = t \,.
 \ee 

  The initial conditions (\ref{icnum}) are  also rescaled in the same manner; 
  they take the same form but in the scaled variables $[\hat{\cdot} ]$. That is the parameters 
  $[A, k_0, l_0, L_X, L_Y]$  are replaced by
\be\label{rescp}
\hat{A} = \eps A, \,\, \hat{k}_0 = \eps^{1/2} k_0, \hat{l}_0 = \eps^{1/2} l_0, \,\,
\hat{L}_X = L_X\, \eps^{-1/2}, \,\,\hat{L}_Y = L_Y\,\eps^{-1/2}  \,.
\ee 
$P_X, P_Y$ are not  rescaled and are chosen as needed for each numerical simulation. 
In these rescaled variables (\ref{rescp}) the initial amplitude should be small, and the initial length scale
should be long. Thus, although there is formally no $\eps $ in (\ref{bls1}) its presence is felt by the choice of 
appropriate  initial conditions. In the original dimensional variables we follow \citet{mg24b, mg25}
to find a suitable estimate of $\eps $, and so choose as a prototype a $5$-second wave
 with frequency $\omega =2\pi/5 = 1.257 s^{-1}$ in a depth of  $h = 10 \,m $.  In the shallow water limit, 
the phase speed $c = (gh)^{1/2}\, m\,s^{-1}$ and so the wavenumber $\kappa = \omega/c = 0.127\, m^{-1}$.
 For the scaling wave amplitude we set $A_0 = 2.5 \, m$, quite large, but well within the dynamical 
 stability range where the crest-trough amplitude ratio to the depth should be less than about $0.75 $. 
 Using the BL scaling (\ref{scale}) these produce values of $\eps = A_0/h = 0.25$, $L = \eps^{-1/2}h = 20.0 \, m$,
 with initial wave amplitude, wavenumber $A =1$, $\kappa_0 =  2.54 $. 
 Then, with the further rescaling (\ref{resc}) to get (\ref{bls1}), these imply that in the 
initial conditions (\ref{icnum}) we set $A \sim 0.25 $ and $\kappa_0 \sim 1.27 $.
We vary $l_0 $ over the range $0.0 - 0.3$ so that $k_0, l_0 $ satisfy KP requirements, and
we choose $L_{X, Y} \sim 30 - 60  $ and note that in the envelopes $ENV(x, y )$ the ratios 
$(x/L_X , y/L_Y ) $ are not changed by (\ref{resc}, \ref{rescp}).

All the numerical results shown here use the form (\ref{bls1}) with $\eps =1 $. 
The results for other smaller values of $\eps $ are easily inferred using the scaling 
(\ref{resc}, \ref{rescp}). As expressed above, the main aims here are to reduce the time 
for nonlinear terms to take effect, and to remove the explicit dependence on $\eps $. 
A crude simple estimate of the nonlinear terms in (\ref{bls1}) is to suppose that  
$f, g, (x, y)$ have  magnitudes $F, G, L$ with  a nonlinear time scale $T$. 
Then, since at least initially the  solution will evolve with speed unity, $f_t = -g$ implies that $F \sim GL $. 
The nonlinear terms are estimated  to have a magnitude $FG/L^2 \sim G^2/L $.  
These balance $g_t $ when $T \sim L/G$, which with $L \sim 2\pi /\kappa_0 \sim 5 $, 
$G \sim 0.25$ gives $T \sim 20 $ in rough agreement with the initial nonlinear deformation 
in the numerical simulations. When $\eps $ is restored and the system is (\ref{bls}) the 
analogous time scale is an order larger, $\eps^{-1}$. 

 Using the typical values $A =0.25, \kappa_0 =1.27 $ as specified above, the simulations 
with envelopes $ENV(x, y )$ tended to produce a strong radiation field in the domain $0 < x < t $
with very little sign  of significant solitary wave formation in $ x > t$. This is very different from analogous
 KdV, KP simulations using the same initial conditions, described by \citet{mg24b, mg25}.  
Part of the explanation is because in the BL system (\ref{bls1}) the solution is described essentially by the 
linear two-way long wave equation with rather weak dispersion and nonlinearity, whereas in 
the KdV, KP equations, the time evolution is for a rescaled slow time relative to the moving 
coordinate $x-t$, with strong nonlinearity and high wavenumber dispersion. The linear dispersion relation for the 
phase speed for downstream waves in the regularised BL equation (\ref{bls1})  is $c =  (1 + \kappa^2/3 )^{-1/2} $
and varies only in the range $0 < c < 1$, and there is no explicit part of  the linearised  (\ref{bls1}) 
which implies evolution of $\kappa $; nonlinearity is needed to achieve that. To explore this further 
we consider the linearised BL system  (\ref{bls1}).  For downstream waves in case A
with no envelopes, the  solution is the sinusoidal wave
\be\label{linbl}
f, g = \frac{A}{\omega_0 } \sin{\theta }, A \cos{\theta }, \,\, \theta = 
(k_0 x + l_0 y -\omega_0 t) , \,\, \omega_0 = \frac{\kappa_0}{(1 + \kappa_0^2/3)^{1/2}}.
\ee
With an envelope, a  schematic  one-dimensional analysis for just the right-going wave 
gives in Fourier space,
\be\label{bllinf}
\ma{F}(f)= \ma{F}(ENV)(k -k_0)\exp{(ikx -i\omega t )}, \,\, \omega = \frac{k}{[1 + k^2/3]^{1/2}}, \,\,
\ma{F}(ENV)(k) = \pi\,L\, \hbox{sech}(\pi k L /2).
\ee  
where to satisfy the initial condition  $k $ in the envelope is shifted by $k_0 $.  
For large times this yields as $t \to \infty $, 
\be\label{bllinsol}
 f= C_0 \frac{\ma{F}(ENV)(k -k_0)}{|\omega_{kk}|^{1/2}}\sin{(kx -\omega t + \theta_0 )}, \,\, 
\frac{x}{t} =c_g= \frac{1}{(1 + k^2/3)^{3/2}}. \ee
 $C_0, \theta_0 $ are constants whose exact values are not needed here. 
The evolving solution  lies in $0 < x < t$  moving with a speed $0< c_g (k) < 1$
where $k$ is a function of  $x/t $. To consider the tendency to steepen and then form 
solitary waves we examine the limit $x/t \to 1$ when $k \to 0$ as solitary waves must 
lie in $x/t > 1$. In this limit $\ma{F}(ENV)(k- k_0) \to \pi\,L\, \hbox{sech}(\pi k_0 L /2)$ and is
  quite small for our parameters. However, $|\omega_{kk}| \to 0 $ as $k \to 0$ and so 
  this linearised solution amplitude increases without bound indicating that solitary waves 
  will eventually form albeit on a long time scale noting that the singularity as $k \to 0 $ in 
  $|\omega_{kk}|^{-1/2}$ must overcome the exponentially small term 
  $\pi\,L\, \hbox{sech}(\pi k_0 L /2)$ as $k_0 L \gg 1$. \textit{Inter alia} this indicates the 
  important role that the envelopes play in the initial amplitude. 

Various schemes are available to overcome these difficulties. \citet{ac11} transformed to the 
travelling coordinate $x- t$, with slow time and transverse dependence, as in the KP equation. 
\citet{mt99} proposed using a high wavenumber filter. Here we rely on the rescaling
 (\ref{resc}, \ref{rescp}) to produce  (\ref{bls1}) and if needed artificially increase the initial amplitude
 to seek results in a reasonable computational time, noting that the physical limit of $0.75 $ 
 described above is not  {\it per se} in the BL equation. We varied the available parameters 
  $[A, k_0, l_0, L_X, L_Y, P_X, P_Y ]$ using mainly the initial conditions cases A, B in  (\ref{icnum}),
  and only show here a small representative sample, set out in table \ref{tab}.  All cases shown here
  have $\eps = 1$, and are without a smoothing function $SMO $, the notation $\text{spg}(x, y) $
  indicates the presence or otherwise of sponge layers and $x_{\text{sh}}, y_{\text{sh}}$ indicate 
  the central location of the initial condition in the event it has been shifted from $(0, 0 )$.
  The number of Fourier modes in the $x, y$ directions are $N_x=2^{11}, N_y=2^{11}$ 
  for cases wide in the $y$-direction,  and $N_x=2^{12}, N_y=2^{7}$ for narrow cases. 
  The time step $\Delta t = 0.1$. 
  
  \begin{table}[h!]
	\centering
	\begin{tabular}{cccccccccc}
		\hline
		 Figures & Case & $A$ & $k_0$ & $l_0$ & $L_X$, $L_Y$ & $\text{spg}(x,y)$ 
		& $x_{\text{sh}}, y_{\text{sh}}$ & $P_X, P_Y$ & $\beta^{\prime}$  \\ \hline
		\ref{fig:zk} & A & 1.0 & 0.1 & 0 & -, - & n, n & -, - & $\pi/k_0$, $\pi$ & 0  \\
		\hline
		\ref{fig:initial}(a), \ref{fig:caseA_wide} & A & 0.5 & 0.2 & 0.2 & 30, 30  & y, y & -300, -300 & $40\pi/k_0$, $40\pi/l_0$ & 0 \\
		\hline
		\ref{fig:initial}(b), \ref{fig:caseB_wide} & B & 0.5 & 0.2 & 0.2 & 30, 30  & y, y & -100, -100 & $40\pi/k_0$, $40\pi/l_0$ & 0  \\
		\hline
		\ref{fig:caseA_narrow} & A & 0.5 & 1.27 & 0.1 & 30, 30  & y, n & -400, 0 & $400\pi/k_0$, $10\pi$ & 0 \\
		\hline
		\ref{fig:caseB_narrow} & B & 0.5 & 1.27 & 0.1 & 30, 30  & y, n & -400, 0 & $400\pi/k_0$, $10\pi$ & 0 \\
		\hline
		\ref{fig:caseAB_narrow}(a) & A & 0.5 & 1.23 & 0.3 & 60, 60  & y, n & -500, 0 & $400\pi/k_0$, $30\pi$ & 0  \\
		\hline
		\ref{fig:caseAB_narrow}(b) & B & 0.5 & 1.23 & 0.3 & 60, 60 & y, n & -500, 0 & $400\pi/k_0$, $30\pi$ & 0 \\
		\hline
		\ref{fig:caseAcir}(a) & A & 0.5 & 0.1 & 0.1 & 30, 30 & y, y & -300, -300 & $20\pi/k_0$, $20\pi/l_0$ & 0 \\
		\hline
		\ref{fig:caseAcir}(b) & A & 0.6 & 0.9 & 0.9 & 30, 30 & y, y & -200, -200 & $90\pi/k_0$, $90\pi/l_0$ & 0 \\
		\hline
		\ref{fig:max}(a) & A & 0.5 & 1.27 & 0.1 & 30, 30  & y, n & -400, 0 & $400\pi/k_0$, $10\pi$ & 0.001  \\
		\hline
		
	\end{tabular}
	\caption{The initial conditions and parameters for each figure; all cases $\eps = 1$. }
	\label{tab}
\end{table}

Figure \ref{fig:zk} is for case A with $k_0=0.1, l_0=0, A=1$ is when the initial condition is just a single 
sinusoid oriented in the $x$-direction.  The aim is to reproduce the formation of a solitary wave train as
found in the pioneering work by \citet{zk65} in simulations of the KdV equation. 
The similarity is astonishingly good, given that the BL equation is not integrable as far as is known,
 unlike the KdV  equation.  Of course, here as time increases we expect solitary wave collisions, 
 but these will presumably be non-conservative. }\\

\begin{figure}[h]
	\centering
	\begin{tabular}{cc}
		\includegraphics[width=0.5\textwidth]{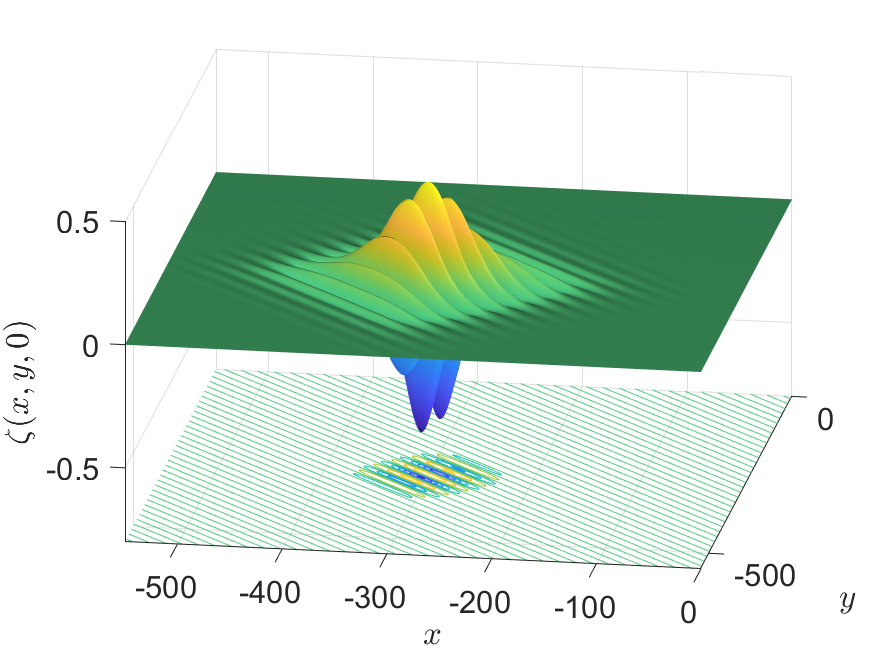} &
		\includegraphics[width=0.5\textwidth]{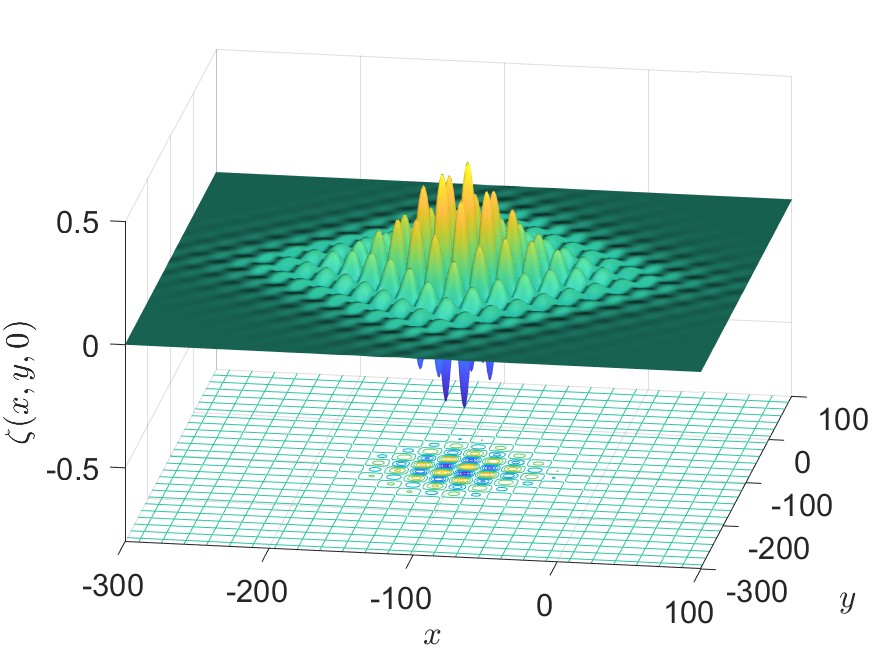} \\
		\small (a)  Case A  & \small (b)  Case B \\
	\end{tabular}
	\caption{Plots of  $\zeta$ at  the initial time, $t=0$, for $k_0=0.2$, $l_0=0.2$, $A=0.5$ for 
	(a) case A, (b) case B.}
	\label{fig:initial}
\end{figure}

Figure \ref{fig:initial} shows the initial value of $\zeta$ found from (\ref{bls1}) for the two cases 
A (a) and  B (b), when $k_0=0.2, l_0=0.2,  A=0.5 $. The centre of the envelope is shifted to $(-300,-300)$ 
for case A  and $(-100,100)$ for case B, while $L_X=L_Y=30$.  \\

\begin{figure}[h]
	\centering
	\begin{tabular}{cc}
		\includegraphics[width=0.5\textwidth]{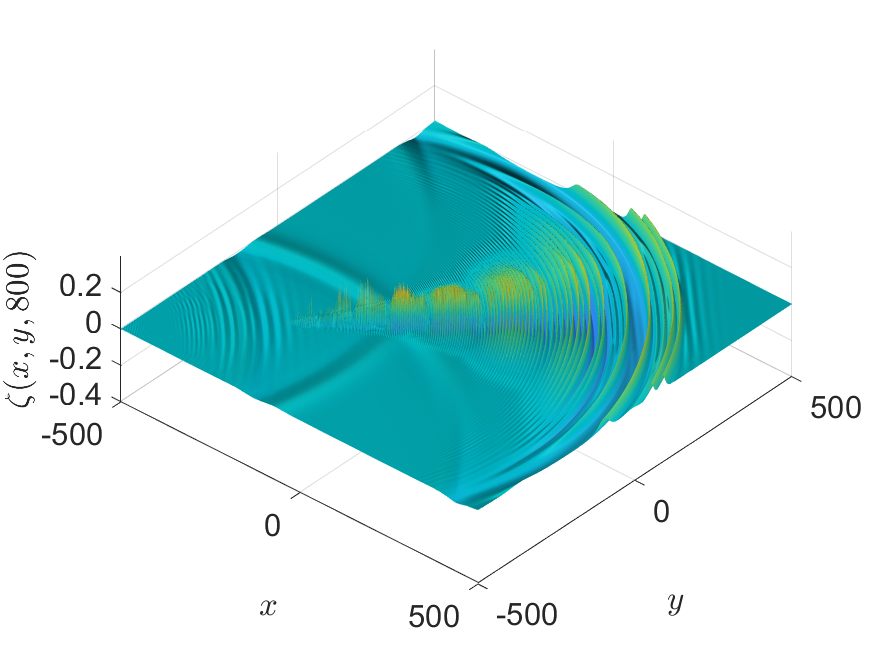} &
		\includegraphics[width=0.5\textwidth]{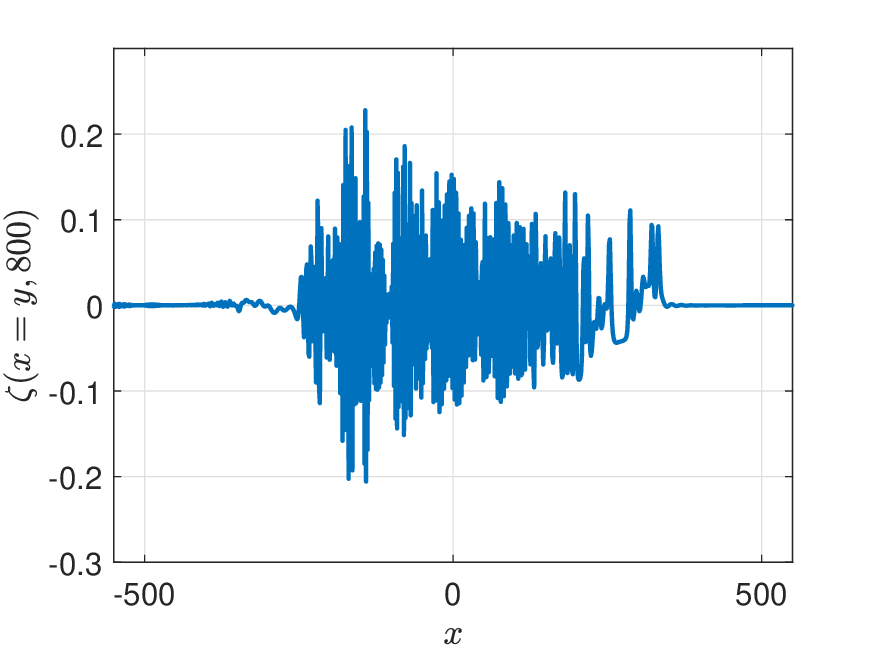} \\
		\small (a)  Surface plot of $\zeta$ & \small (b) Plot of $\zeta$ along $x=y$ \\
	\end{tabular}
	\caption{Case A (wide): Visualisation of $\zeta$ at $t=800$ for $k_0=0.2$, $l_0=0.2$, $A=0.5$,
	 (a)  surface plot, (b) plot along $x=y$.}
	\label{fig:caseA_wide}
\end{figure}

Figure \ref{fig:caseA_wide} (a) shows a plot of $\zeta$ at $t=800$ for a wide configuration 
 when the initial condition is case A shown in  figure \ref{fig:initial} (a). Radiation propagates 
 upstream and circular solitary waves start forming downstream after a large time. 
A plot of $\zeta$ along the line $x=y$  is shown  in figure \ref{fig:caseA_wide} (b), 
small solitary waves start to form at the extreme right-hand side. 
A much longer time is needed to  see their full development. \\

\begin{figure}[h]
	\centering
	\begin{tabular}{cc}
		\includegraphics[width=0.5\textwidth]{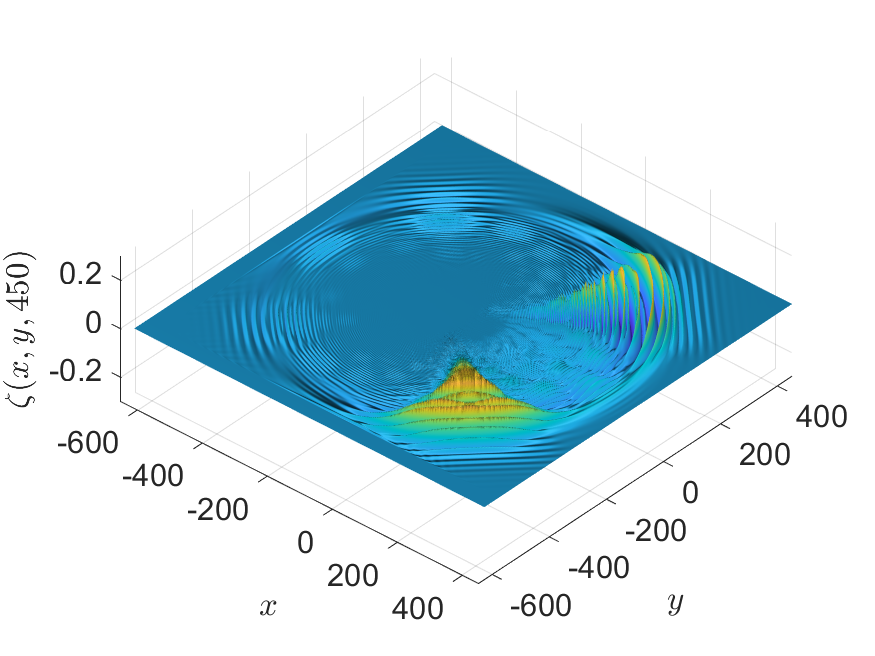} &
		\includegraphics[width=0.5\textwidth]{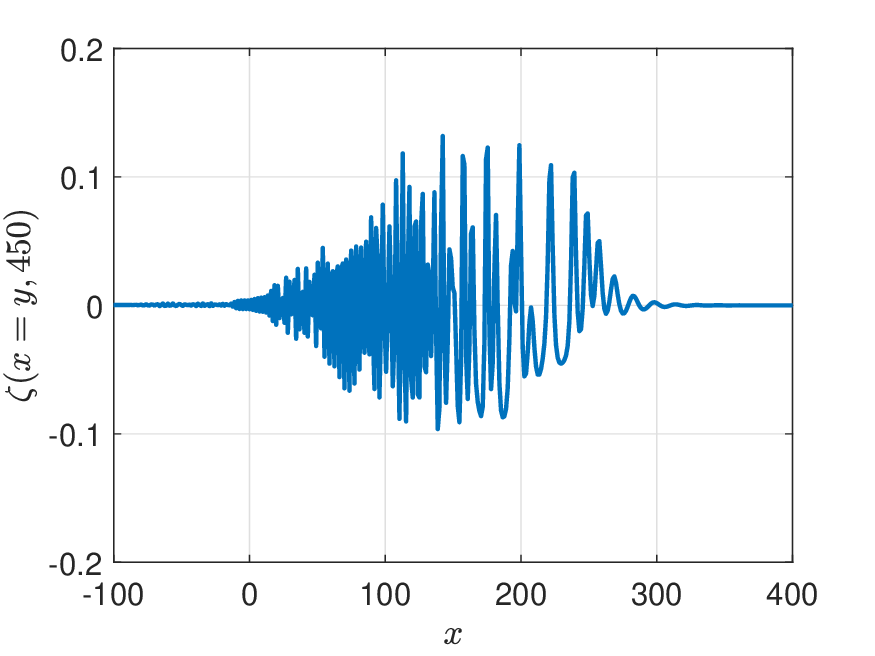} \\
		\small (a)  Surface plot of $\zeta$  & \small (b) Plot of $\zeta$ along $x=y$ \\
	\end{tabular}
	\caption{Case B (wide): Visualisation of $\zeta $ at $t=450$ for $k_0=0.2$, $l_0=0.2$, $A=0.5$. 
	(a)   Surface plot, (b) Plot along $x=y$.}
	\label{fig:caseB_wide}
\end{figure}

Figure \ref{fig:caseB_wide} (a) shows the   surface  plot of $\zeta$ at $t=450$ for a wide configuration. 
when the initial condition is case B shown in figure \ref{fig:initial} (b).  At an  early time in the simulation, 
 the initial envelope is split into two wave groups,  propagating along the lines $x=y$ and $x=-y$. 
 We see clearly  circular waves developing due to a combination of nonlinearity and 
 dispersion where the $y$-dependence is crucial here as the BL system (\ref{bls1}) is isotropic. 
 A plot of $\zeta$ along the line $x=y$  is shown Figure \ref{fig:caseB_wide} (b);
 again small solitary waves start to form at the extreme right-hand side.
 A much longer time is needed to  see their full development and in this case the simulation 
 was stopped at $t=450$. \\

\begin{figure}[h]
	\centering
	\begin{tabular}{cc}
		\includegraphics[width=0.5\textwidth]{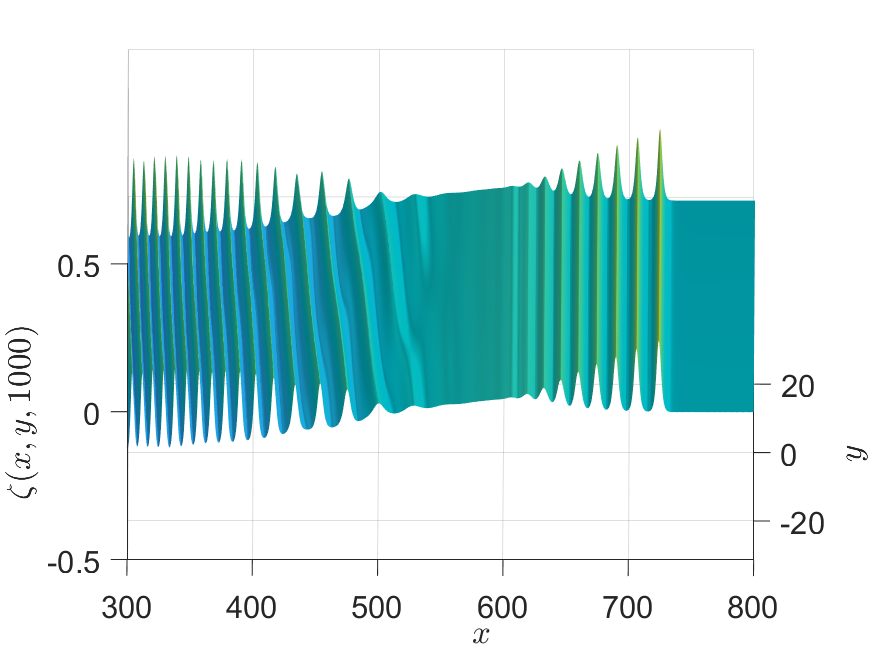} &
		\includegraphics[width=0.5\textwidth]{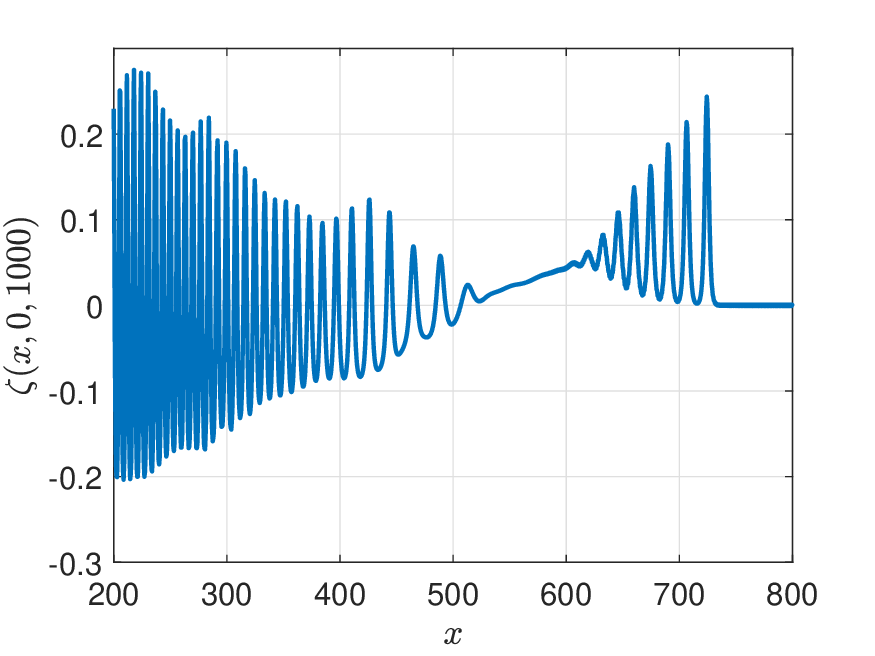} \\
		\small (a) Surface plot of $\zeta$ & \small (b) Plot of $\zeta$ along $y=0$ \\
	\end{tabular}
	\caption{Case A (narrow): Visualization of $\zeta$ at $t=1000$ for $k_0=1.27$, $l_0=0.1$, $A=0.5$. 
	(a)   Surface plot, (b) Plot along $y=0$.}
	\label{fig:caseA_narrow}
\end{figure}

 Figure \ref{fig:caseA_narrow} (a) shows the   surface plot of $\zeta$ at $t=1000$ for a narrow 
configuration, where we choose  a small  $P_Y$ and a large $P_X$. 
The initial condition is case A with $k_0=1.27, l_0=0.1$, $A=0.5$. The main flow is now in the
$x$-direction, so sponge layers are applied at the far field of the $x$-domain,
 while we have fully periodic boundary conditions in the $y$-domain. There are some wave 
 reflections from the $y$-boundaries and these waves enhance the formation of stable 
 solitary waves, leading to a soliton gas in a  larger time simulation. The train of solitary waves 
 is clearly seen in figure \ref{fig:caseA_narrow} (b) around $x=600-700$. This case is similar 
 to our previous work on the  KPB equation, \citet{mg25}. \\

\begin{figure}[h]
	\centering
	\begin{tabular}{cc}
		\includegraphics[width=0.5\textwidth]{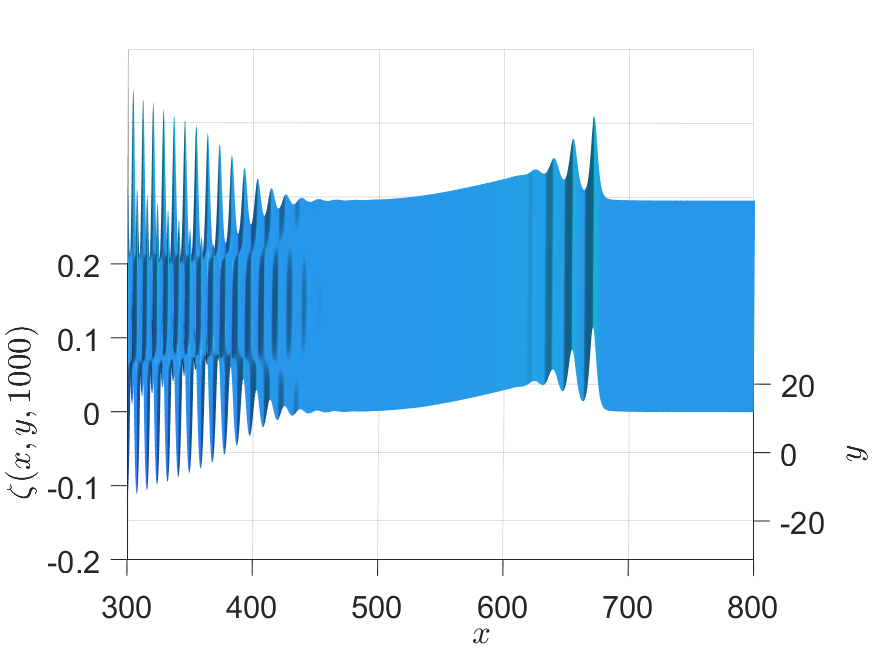} &
		\includegraphics[width=0.5\textwidth]{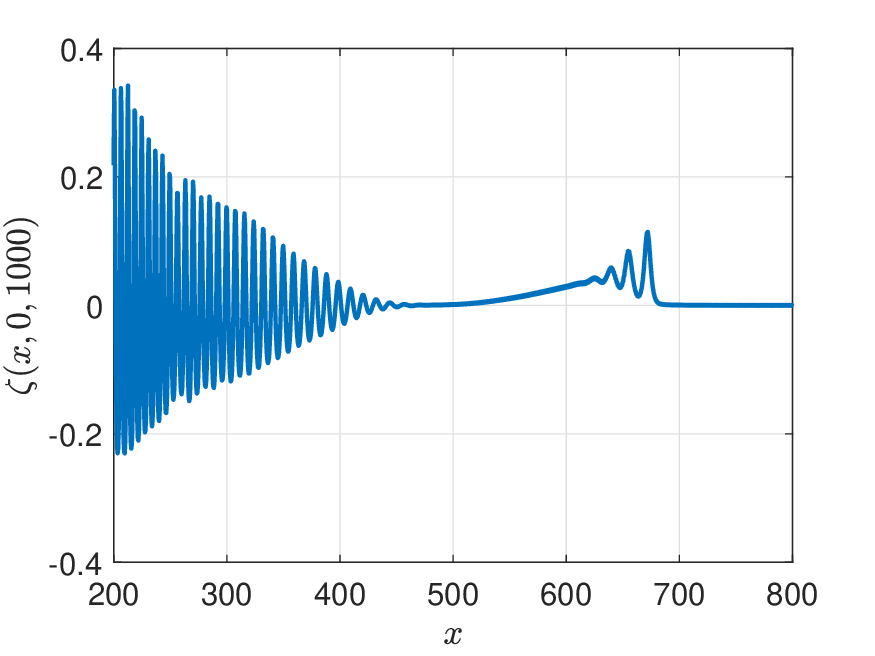} \\
		\small (a)   Surface  plot of $\zeta$ & \small (b) Plot of $\zeta$ along $y=0$ \\
	\end{tabular}
	\caption{Case B (narrow): Visualisation of $\zeta$ at $t=1000$ for $k_0=1.27$, $l_0=0.1$,
	$A=0.5$. (a)   Surface plot, (b) Plot along $y=0$.}
	\label{fig:caseB_narrow}
\end{figure}

Figure \ref{fig:caseB_narrow} (a) shows the   surface plot of $\zeta$ at $t=1000$ for a 
narrow configuration. The initial condition is case B with $k_0=1.27, l_0=0.1$,  $A=0.5$, 
the same as in Figure \ref{fig:caseA_narrow}.  Solitary waves start to form at a large time 
around $x=600-700$  as seen in Figure \ref{fig:caseB_narrow} (b), but the amplitude is smaller 
than those in case A.  Again this case is similar to our previous work on  the KPB equation, \citet{mg25}. \\

\begin{figure}[h]
	\centering
	\begin{tabular}{cc}
		\includegraphics[width=0.5\textwidth]{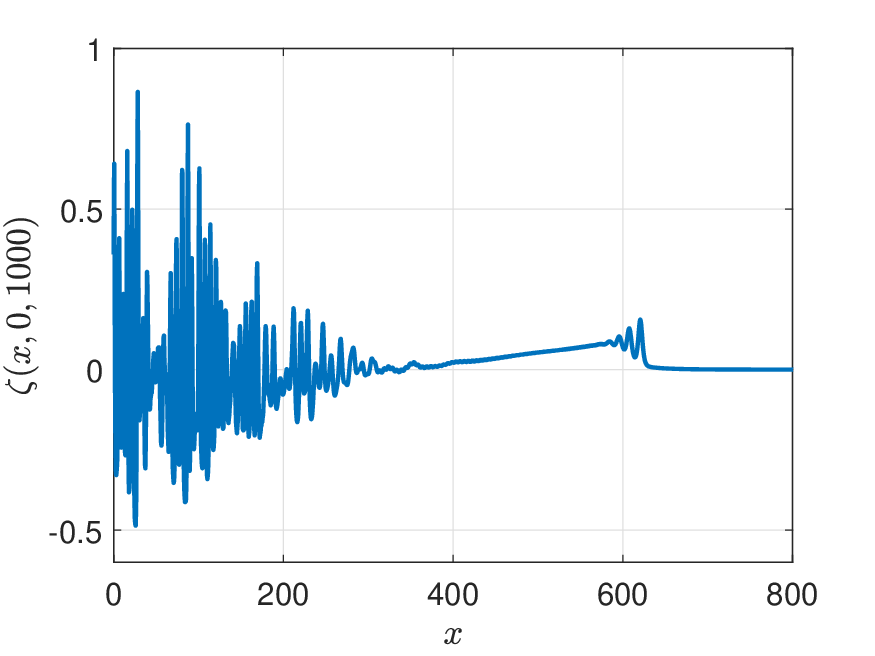} &
		\includegraphics[width=0.5\textwidth]{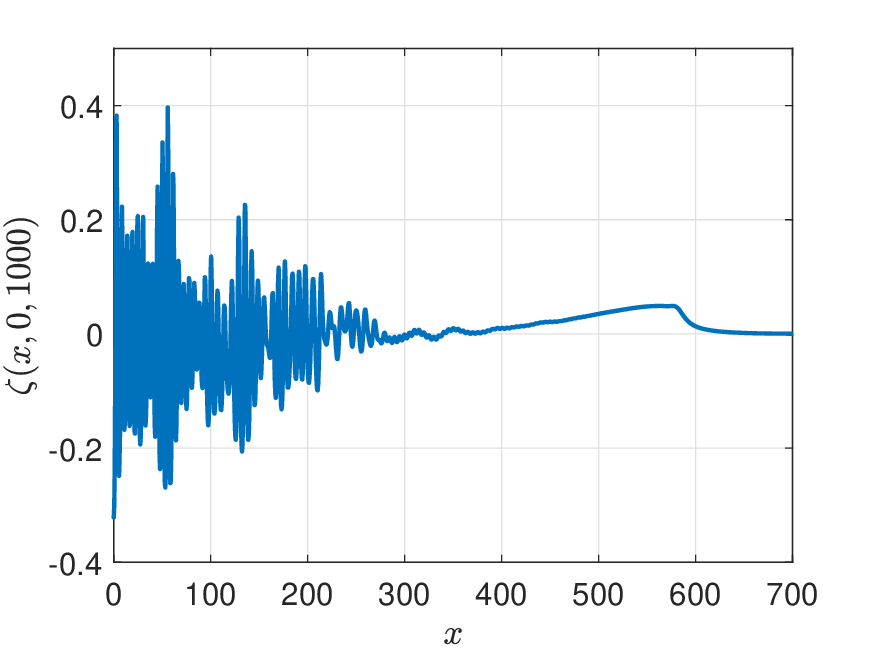} \\
		\small (a) Case A & \small (b) Case B \\
	\end{tabular}
	\caption{Plots of $\zeta$ along $y=0$ at $t=1000$ for two cases with $k_0=1.23$, $l_0=0.3$, 
	and $A=0.5$. (a) Case A (narrow), (b) Case B (narrow).}
	\label{fig:caseAB_narrow}
\end{figure}

Figure \ref{fig:caseAB_narrow} shows plots of $\zeta$ along $y=0$ for cases A and B when 
$k_0=1.23, l_0=0.3$, $A=0.5$. These have a larger $l_0 $ than above, but here we 
show only a narrow case. Solitary waves  start to form in case A, but need a very large 
time to form in case B. \\

\begin{figure}[h]
	\centering
	\begin{tabular}{cc}
		\includegraphics[width=0.5\textwidth]{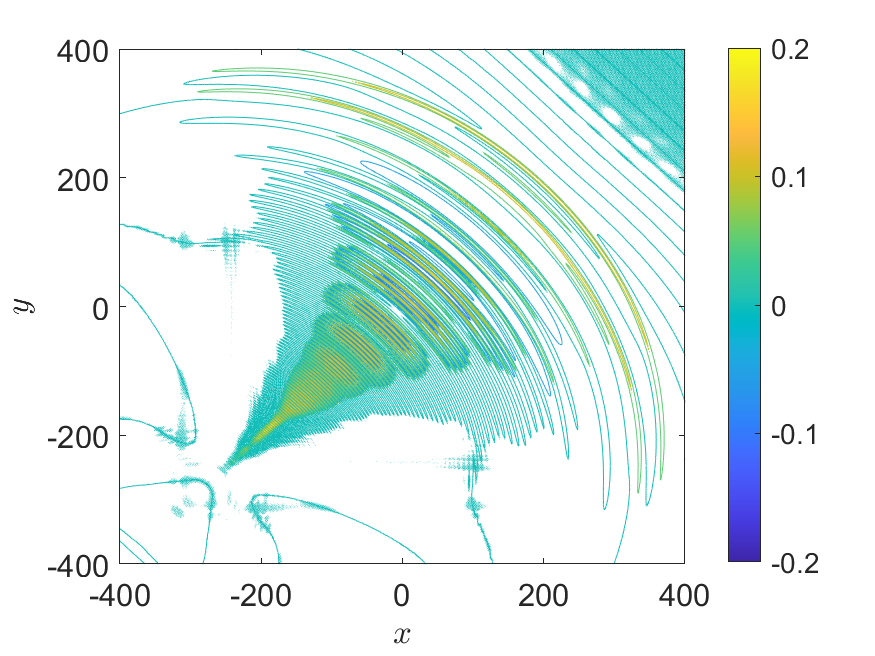} &
		\includegraphics[width=0.5\textwidth]{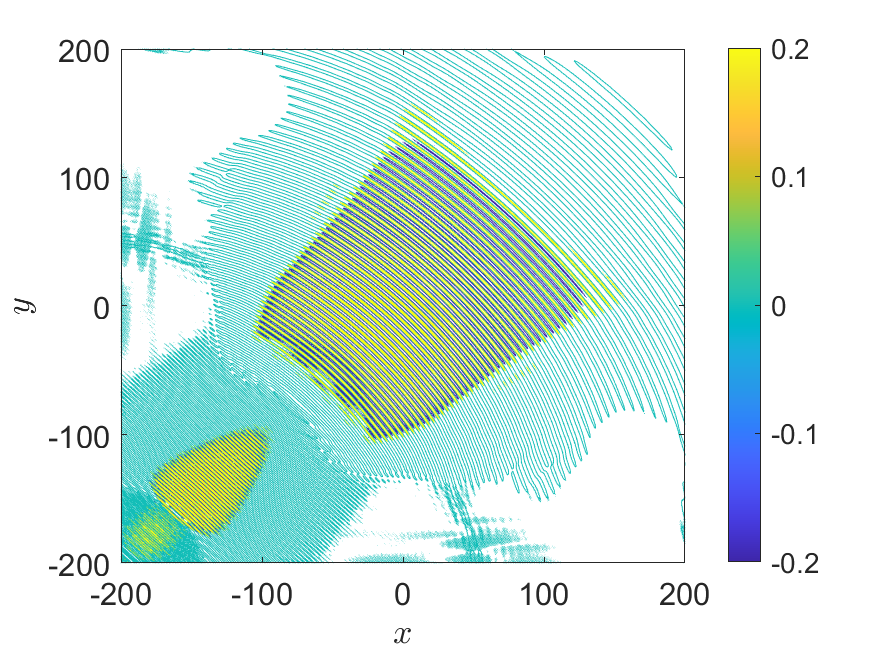} \\
		\small (a) $k_0=l_0=0.1$ & \small (b) $k_0=l_0=0.9$  \\
	\end{tabular}
	\caption{Case A  (wide): Contour plots of $\zeta$ at $t=600$. (a) $k_0=l_0=0.1, A=0.5$, 
	(b) $k_0=l_0=0.9, A=0.6$ }
	\label{fig:caseAcir}
\end{figure}

Figure \ref{fig:caseAcir} shows case A for  (a) $k_0=l_0=0.1$, $A = 0.5$ and  (b) 
$k_0=l_0=0.9 $ $A=0.6$ at $t=600$ in order to exhibit how two-dimensional spatial dispersion 
forms a curved wave front, in this case circular. This  dispersive effect is more marked in case 
 (a) for small $\kappa_0 $ than in case (b) with a larger $\kappa_0$. 
There is no sign of solitary wave train formation in case (a) at a finite time, but possibly  that may be 
present in the leading waves of very small amplitude in case (b). This isotropic behaviour in the 
BL system (\ref{bls1}) cannot be found in the KP equation due to the relative uppression there of 
dispersion in the $y$-direction. We infer that for a solitary wave train to form in the BL equation 
requires a larger initial amplitude $A$ and small initial wavenumber $\kappa_0 $.\\

\begin{figure}[h]
	\centering
	\begin{tabular}{cc}
		\includegraphics[width=0.5\textwidth]{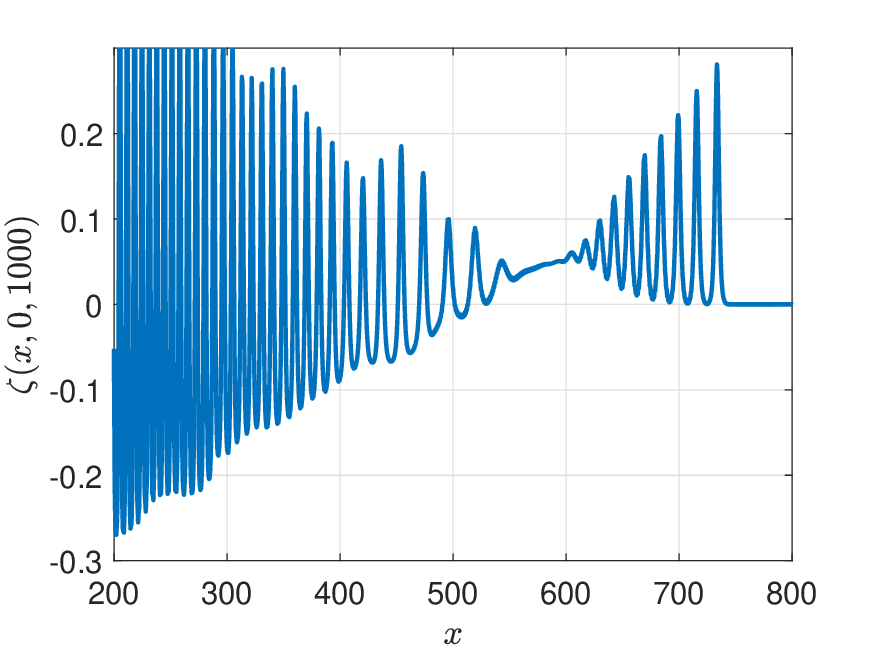} &
		\includegraphics[width=0.5\textwidth]{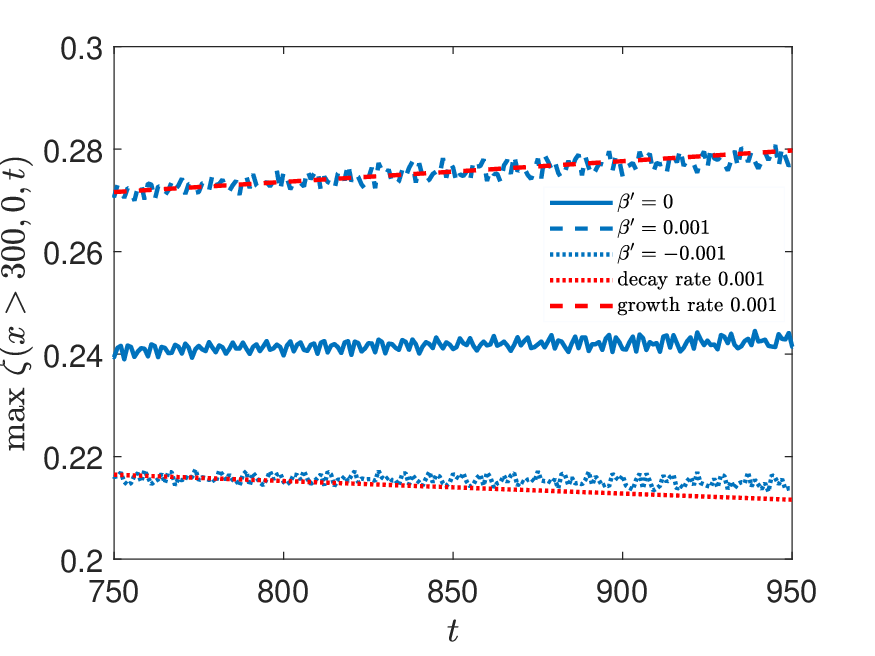} \\
		\small (a) Plot of $\zeta$ along $y=0$ & \small (b) Maximum of $\zeta$  \\
	\end{tabular}
	\caption{Case A (narrow): $k_0=1.27, l_0=0.1, A=0.5$. (a) Plot along $y=0$ at $t=1000$, 
	$\beta^{\prime} = 0.001$.  (b) Maximum of $\zeta(x>300,0,750<t<950)$ with predicted 
	growth and decay rates when $\beta^{\prime}=\pm 0.001$. }
	\label{fig:max}
\end{figure}

Figure \ref{fig:max} (a) has  the same parameter and initial value settings as \ref{fig:caseA_narrow}(b)
but for $\beta^{\prime} =0.001 > 0$. This was  a narrow case A for $k_0=1.27, l_0=0.1$, $A=0.5$ and showed significant formation of a  solitary wave train. With $\beta^{\prime} > 0 $ both the downstream solitary waves 
and the upstream radiation increase in amplitude, see the plots in figure  \ref{fig:max} (a).
The maximum of $\zeta$ is tracked  when $750 < t <950$, for $x>300$ and $y=0$. 
The  solitary wave amplitude downstream is compared with the estimate (\ref{Gt}) from the
modulation theory of subsection \ref{swt} in figure  \ref{fig:max} (b), with good agreement as we found for the 
KPB equation in \citet{mg25}.  Analogous outcomes hold for  other parameter settings, 
for both $\beta^{\prime}>0, <0 $,  but are not shown here. \\
  
\section{Summary and discussion}
\label{summary}

In this paper we have examined the evolution of wind-generated shallow water waves in the framework of 
a modified regularised  Benney-Luke (BL) equation. The wind forcing modification is based on Miles critical level instability mechanism, \citet{mi57}, and is analogous to our previous work on the KdV and KPB equations, \cite{mg24b, mg25}. A key essential difference is that the BL equation is isotropic in the two horizontal spatial variables, $x, y $, whereas the KdVB equation is uni-directional and the KPB equation is expressed in a 
reference frame moving in the wind direction,  $x$, with weak dispersion in the transverse $y$-direction. 
The asymptotic derivation from the full Euler equations is described in  section \ref{bleq} and based on the 
original work of \citet{bl64}. It is based on two small parameters, $\eps $ the ratio of wave amplitude to the undisturbed  water depth, and $\beta $ measuring long-wave dispersion, see (\ref{scale}). 
Since we expect to see the generation of solitary waves we make the usual assumption that 
$\eps = \beta$ expressing the balance between weak nonlinearity and long-wave dispersion.  
The wind forcing term contains a modified Miles parameter $\beta^{\prime}$
where positive/negative  values lead to wave growth/decay  respectively.

In subsection \ref{cn} we recover the cnoidal and solitary wave solutions found by \citet{bl64} and in 
subsection \ref{mod} we describe the asymptotic theory of wave modulations with a main focus on 
the solitary wave train limit in subsection \ref{swt}.  This theory closely follows the analogous theory 
for the KPB equation in \citet{mg25}, as the travelling wave solutions of the BL equation are very close
 to those of the KP equation. This modulation analysis is used to give explicit estimates of wave growth/decay according as $\beta^{\prime } $ is positive/negative respectively.
 
The main part of this paper is section \ref{num} where we present some numerical results. 
Simulations of the BL equation (\ref{bls}) are performed using a pseudo-spectral method in space 
and a fourth-order Runge-Kutta method in time. The code is described in an Appendix, and was tested 
using the travelling wave solutions of section \ref{tw}.  The initial conditions are sinusoidal waves enclosed in a slowly varying envelope designed to produce wave packets. Two main cases are considered, (\ref{icnum}): 
Case A with one sinusoidal component, and Case B with two sinusoidal components, 
both characterised by $x, y$ wavenumbers $k_0, l_0 $ and an amplitude $A$.  The envelopes 
ensure smooth decay at the boundaries, periodic in the numerical simulations.  
The wind forcing is in the $x$-direction, and we  consider two cases when the  periodic domain
 is relatively narrow or wide in the $y$-direction. 
 
 Many simulations were performed, and we present here only a small representative sample. 
Our focus is on  solitary wave trains and we find that they are eventually formed over large time scales 
due to the weak nonlinearity and weak dispersion in the modified BL equation. With no wind forcing, 
in the narrow domain case we find that small initial wavenumbers and  a large initial amplitude 
provide the most favourable conditions for the formation of solitary waves.  When wind forcing is included,
in the narrow domain case the wave amplitudes grow/decay in agreement with the predictions from 
the modulation theory, similar to our study of the KPB equation in \citet{mg25}. But in the wide 
domain case, $y$-dependent dispersion dominates leading to curved wave fronts and delaying, 
even preventing, the formation of solitary waves.

Although the BL equation is isotropic in the horizontal spatial variables $x, y$, a property
 inherited from the full Euler equations, it is limited numerically because at leading order  it is 
 the linear two-dimensional long  wave equation and, as described in section \ref{num}, it takes a long time, 
 of order $\eps^{-1}$, for nonlinearity and dispersion to take effect and then produce solitary waves. 
 Unlike the asymptotic reductions to KdV, KP models which evolve on a slow time scale relative to the moving coordinate $x-t$, in a BL model the solution must numerically first follow this moving coordinate 
 before evolving.  Another feature here is that the BL equation we study is regularised, so that 
 there is no spurious high wavenumber instability. This is absolutely necessary for numerical 
 simulations but has the effect here that  the high wavenumber dispersion needed to prevent 
 nonlinear steepening is somewhat inhibited.  Both these features persist even when 
 we use the rescaled form (\ref{bls1}) where formally $\eps =1$.  Nevertheless, the 
 regularised BL equation, with or without wind forcing, is a useful tool to study two 
 space-dimensional shallow water waves. Possible improvements and extensions
 include going to the next $\eps $ term in the BL derivation.
 
Operational wind wave forecasting is typically based on the Hasselmann equations 
for the wave spectrum, and were developed essentially for deep-water waves.  
Although operational adjustments can be and are made, they are not totally suitable for 
wind waves in shallow water, where the evidence from works such as \citet{os14} and
 our work in this paper and in \citet{mg24b, mg25} is that a model describing a soliton gas is needed.
 Whether a model such as this BL system, or a similar Boussinesq-type model, 
can be used for operational forecasts remains a challenge.  We note that \citet{oojr09} 
developed a  kinetic theory of the classical  Boussinesq equations and compared that 
    with the shallow water limit of the  Hasselmann equation, with promising results.  
    Although some issues about resonances in shallow water remain, this would seem to be a 
    promising direction for future research. We suggest that an analogous kinetic theory of the 
     regularised BL equation might be fruitful.

\section*{Appendix: Numerical scheme}

From the BL equation written in the system form (\ref{bls}), 
\be\label{blsnum}
f_t = - g, \quad g_t = -\nabla^2 f  + \eps [\frac{1}{3}  \nabla^2 g_{t}  - g\nabla^2 f  - 2f_x g_{x} -  2f_y g_{y}]
+  \alpha g_{xt} \,.
\ee

For $g_t $ we schematically factorise the operator, 
\be\label{gtf}
g_{t} =   \frac{1}{  (1-\frac{\eps}{3}\nabla^2 - \alpha \partial_x ) }   \{ -\nabla^2 f - \eps \left[g \nabla^2 f +2f_x g_{x} +2f_y g_{y} \right]\}
\ee
Then define
\be\label{odef}
\frac{\partial f}{\partial t} = -g := F_1(f, g, t)
\ee
\be\label{odeg}
\frac{\partial g}{\partial t} = \frac{1}{  (1-\frac{\eps}{3}\nabla^2 - \alpha \partial_x ) }  \left\{ -\nabla^2 f - \eps \left[g\nabla^2 f  + 
2f_x g_x +  2f_y g_y \right]  \right\} := F_2(f, g, t)
\ee
These are two first-order ordinary differential equations  for $f, g$ with initial conditions $f(x,y,0)$ and $g(x,y,0)$. These equations will be solved numerically by the fourth order Runge-Kutta method in time 
pseudo spectral method in space. We assume periodic boundary conditions based on the Fourier basis.
Let $\mathcal{F} (f)$ denotes the two-dimensional Fourier transform of $f$, $\mathcal {F}(f) = \hat{f} (k_x, k_y, t)$ where $k_x$ and $k_y$ are wavenumbers in Fourier space. Similarly, 
$\mathcal {F}(g) = \hat{g} (k_x, k_y, t)$. $\mathcal{F}^{-1} (\hat{g})$ 
is the inverse Fourier transform of $\hat{g}$. 
Then we approximate the derivative and linear operation for each term in (\ref{odeg}) using the Fourier transform, then transforming back to the physical space to apply the product of two terms as follows. \\

\noindent First term: define $\mathcal{F} (\nabla^2) = -(k_{x}^{2} + k_{y}^{2}) = -\kappa^2$,   
$\mathcal{F}  \left(- \frac{\nabla^2}{(1-\frac{\eps}{3}\nabla^2 - \alpha \partial_x ) } f \right)  = \frac{\kappa^2}
 {(1 + \frac{\eps}{3} \kappa^2 - \alpha i k_x) }  \hat{f} $, so 
\[
-\frac{\nabla^2} {(1-\frac{\eps}{3}\nabla^2 - \alpha \partial_x )}  f  = \mathcal{F}^{-1} \left( \frac{\kappa^2}
{(1 + \frac{\eps}{3} \kappa^2 - \alpha i k_x) }  \hat{f}  \right)
\]
Note that the denominator is well behaved, and is not zero for all wavenumbers. \\

\noindent Second term: $ \nabla^2 f = \mathcal{F}^{-1} (-\kappa^2 \hat{f}) := a_1$, let $b_1 = ga_1$, then find
\[
\mathcal{F} \left( -\frac{\eps} {(1-\frac{\eps}{3}\nabla^2 - \alpha \partial_x ) } b_1 \right) = -\frac{\eps}
 { (1 + \frac{\eps}{3} \kappa^2 - \alpha i k_x) }  \hat{b}_1
\] 
where $\hat{b}_1$ is the Fourier transform of the product of $g$ and $a_1$. Then 
\[
-\frac{\eps}{(1-\frac{\eps}{3}\nabla^2 - \alpha \partial_x ) } ( g \nabla^2 f ) = \mathcal{F}^{-1} \left( -\frac{\eps} 
{ (1 + \frac{\eps}{3} \kappa^2 - \alpha i k_x) }  \hat{b}_1 \right)
\] 
\\

\noindent Third term: Let $f_x = \mathcal{F}^{-1} \left( i k_x \hat{f} \right) := a_2$ and $g_x = \mathcal{F}^{-1} \left( i k_x \hat{g} \right) := b_2$
\[
-\frac{2 \eps}  {(1-\frac{\eps}{3}\nabla^2 - \alpha \partial_x )}  (f_x g_x) = \mathcal{F}^{-1} \left( -\frac{2 \eps} {  (1 + \frac{\eps}{3} \kappa^2 - \alpha i k_x) }  \hat{c}_2 \right)
\] 
where $\hat{c}_2$ is the Fourier transform of the product of $a_2$ and $b_2$. \\

\noindent Fourth term: Let $f_y = \mathcal{F}^{-1} \left( i k_y \hat{f} \right) := a_3$ and $g_y = \mathcal{F}^{-1} \left( i k_y \hat{g} \right) := b_3$
\[
-\frac{2 \eps} {(1-\frac{\eps}{3}\nabla^2 - \alpha \partial_x ) } (f_y g_y) = \mathcal{F}^{-1} \left( -\frac{2 \eps}
 { (1 + \frac{\eps}{3} \kappa^2 - \alpha i k_x) }  \hat{c}_3 \right)
\] 
where $\hat{c}_3$ is the Fourier transform of the product of $a_3$ and $b_3$. 

Thus  all terms on the right- hand side of (\ref{odeg}) are evaluated. Then  we solve this 
ordinary differential equation system to obtain $f$ and $g$ at the next time step using the
 fourth-order Runge Kutta method. The surface elevation $\zeta$ is found using (\ref{bls1}). 
The boundary conditions in both $x$ and $y$ domains are periodic in this Fourier basis. 
To absorb radiating waves moving to the far field near the boundaries while maintaining periodicity  
we use a sponge layer technique. The sponge layer function in the $x$ direction is defined by
\be\label{sponge}
\text{spg}(x) = \frac{\nu}{2} \left[ 1+ \tanh(\kappa_s(x - x_m)) + (1-\tanh(\kappa(x+x_m)))  \right]
\ee
where $x_m = w_d P_X \pi/2$, $w_d$ controls the width of sponge layer from the ends of boundaries, 
$\nu$ is the magnitude of the sponge layer, and $\kappa_s$ controls the smoothness. 
The sponge layer in the $y$ direction, $\text{spg}(y)$ is defined similarly. 
The sponge layer in the $x$ direction is applied to $f, g$ in the physical space by 
\be 
f(x,y,t+dt) = f (1 - \text{spg}(x)), \, g(x,y,t+dt) = g (1 - \text{spg}(x)) \,.
\ee
Finally, $f(x,y,t+dt), g(x,y,t+dt)$ are transformed back into the Fourier space again to update $\hat{f}, \hat{g}$ for the next time step. In this work, we usually set $\nu=1, \kappa_s=0.4$ and $w_d=1.85$.

\newpage
\bibliographystyle{apalike} 
\bibliography{ref}

\newpage
\begin{figure}[ht]
	\centering
	\includegraphics[width=12cm]{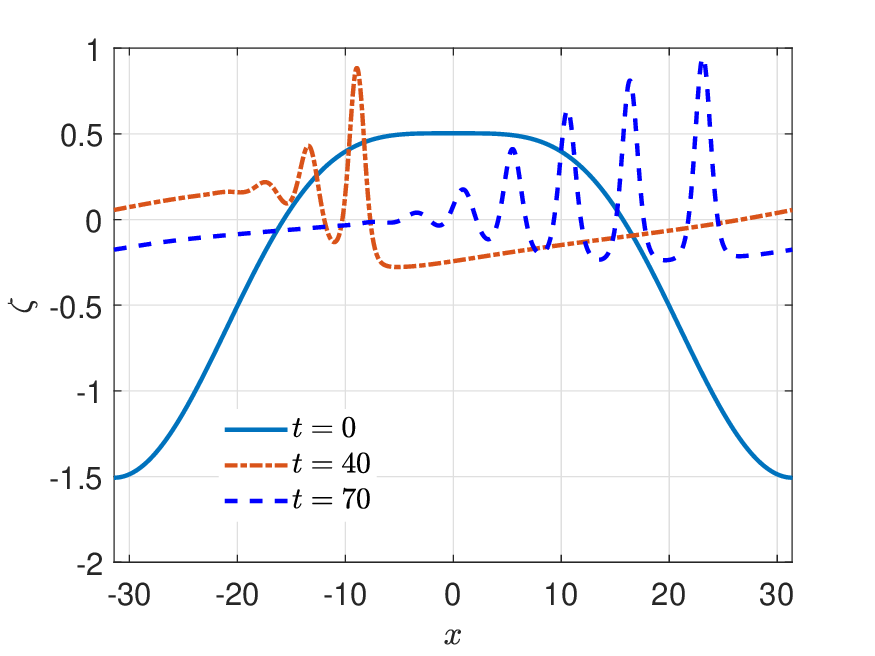} 
	\caption{Case A: Plot of $\zeta$ along $y=0$ at $t=0, 40, 70$, $k_0=0.1, l_0=0, A=1$, 
	without envelopes $ENV(X,Y)$.   }
	\label{fig:zk}
\end{figure}

\end{document}